\documentclass{emulateapj}
\usepackage{natbib}

\newcommand\etal{et al.}
\newcommand\ie{i.e.}
\newcommand\eg{e.g.}
\newcommand\rvec{{\bf r}}
\newcommand\vvec{{\bf v}}
\newcommand\vrad{v_r}
\newcommand\vphi{v_\phi}
\newcommand\erad{{\bf e}_r}
\newcommand\ephi{{\bf e}_\phi}
\newcommand\ez{{\bf e}_z}
\newcommand\nhat{{\bf n}}
\newcommand\vlos{v_{\rm los}}
\newcommand\ri{r_{\rm i}}
\newcommand\ro{r_{\rm o}}
\newcommand\phii{\phi_{\rm i}}
\newcommand\phio{\phi_{\rm o}}
\newcommand\vi{v_{\rm i}}
\newcommand\vki{v_{\rm K,i}}
\newcommand\vo{v_{\rm o}}
\newcommand\vko{v_{\rm K,o}}
\newcommand\emin{E_{\rm min}}
\newcommand\romin{r_{\rm o, min}}
\newcommand\vlosmax{v_{\rm los, max}}

\newcommand\cf{cf.}
\newcommand\kms{\ifmmode{\rm km\ s^{-1}}\else$\rm km\ s^{-1}$\fi}
\newcommand\sound{c_{\rm s}}
\newcommand\rhoe{\rho_{\rm e}}
\newcommand\gvec{{\bf g}}
\newcommand\avec{{\bf a}}
\newcommand\cdrag{C_{\rm D}}
\newcommand\reynolds{{\rm Re}}
\def\eps@scaling{1.0}
\newcommand\plotthree[3]{{
 \typeout{Plotthree included the files #1 #2 #3}
 \centering
 \leavevmode
 \columnwidth=.31\columnwidth
 \includegraphics[width={\eps@scaling\columnwidth}]{#1}
 \includegraphics[width={\eps@scaling\columnwidth}]{#2}
 \includegraphics[width={\eps@scaling\columnwidth}]{#3}
}}

\shorttitle{M86 FROM CHANDRA}
\shortauthors{RANDALL ET AL.}
\slugcomment{Astrophysical Journal, submitted}

\begin{document}

\title{Chandra's View of the Ram Pressure Stripped Galaxy M86}

\author{S.\ Randall\altaffilmark{1}, P.\ Nulsen\altaffilmark{1}, W.\ R.\ Forman\altaffilmark{1}, C.\ Jones\altaffilmark{1}, M.\
  Machacek\altaffilmark{1}, S.\ S.\ Murray\altaffilmark{1} and B. Maughan\altaffilmark{2}
}

\altaffiltext{1}{Harvard-Smithsonian Center for Astrophysics, 60
  Garden St., Cambridge, MA 02138, USA}
\altaffiltext{2}{Astrophysics Group, Bristol University, Tyndall
  Avenue, Bristol, BS8 1TL, UK}

\begin{abstract}
We present results from a mosaic of nine {\it Chandra} observations of
M86 and the surrounding field.  We detect three main diffuse
components: the Virgo ICM at $\sim$2.4~keV, the extended halo of M86 at
$\sim$1.2~keV, and the cooler central and stripped gas of M86 at
$\sim$0.8~keV.  The most
striking feature is a long tail of emission, which consists of
a plume $\sim$ 4\arcmin north of M86 and two main extensions emanating
from the plume.
Based on the morphology and temperature structure of the tail, we
conclude that it is formed 
by ram pressure stripping of M86 as it falls into
the Virgo cluster and 
interacts with the Virgo ICM, in agreement with earlier work.  
The tail is 150~kpc in projection,
and a simple estimate gives a lower limit on the true length of the
tail of 380~kpc, making this the longest ram
pressure stripped tail presently known.  
The total gas mass in the plume ($\sim 7 \times 10^8 \, {\rm
  M_\odot}$) and tail ($\sim 1 \times 10^9 \, {\rm
  M_\odot}$) is about three times that in the core of M86, which
supports the scenario where most of the gas was stripped rapidly and recently.
The projected position of the 
plume can be understood if M86 has an aspherical potential, as
suggested by optical isophotes.  Ram pressure stripping from an
aspherical potential can also explain the split ``double tails'' seen
in M86 and in other Virgo cluster galaxies in the field.
The large line-of-sight
velocity of M86 (1550~\kms with respect to M87), its position 
relative to the Virgo cluster, and the orientation of the tail tightly constrain its orbital
parameters.  
The data are inconsistent with a radial orbit, and imply
inner and outer turning radii of $\ri \approx 300$~kpc and $\ro \ga
8.8$~Mpc, indicating that M86 is, at best, only weakly bound to the Virgo cluster.
\end{abstract}


\section{Introduction} \label{sec:intro}

M86 (NGC~4406) is a bright elliptical (E3/S0) galaxy in the Virgo
cluster of galaxies.  It is the dominant member of one of the larger
subgroups within Virgo (Binggeli \etal\ 1993; B\"ohringer \etal\ 1994;
Schindler \etal\ 1999).
Its line-of-sight velocity relative to M87, the dominant member of the
Virgo cluster, is -1550~km~s$^{-1}$, much higher than the average
cluster velocity dispersion (Smith et al.\ 2000).  The X-ray surface brightness
distribution is unusual, with a large ``plume'' extending to the
northwest from M86, which was first noticed as part of a survey of
Virgo cluster
galaxies undertaken with the {\it Einstein Observatory} (Forman et al. 1979).
The galaxy has an optical asymmetry that extends in a direction
similar to the direction of the plume (Nulsen \& Carter 1987; Mihos
\etal\ 2005).
Several authors have interpreted this plume as arising from ram
pressure stripping due to strong interactions with the Virgo ICM
(Forman \etal\ 1979; Fabian \etal\ 1980; Takeda \etal\ 1984; Knapp
\etal\ 1989; Bregman \&
Roberts 1990; White \etal\ 1991; Rangarajan \etal\ 1995).  Elmegreen
\etal\ (2000) find dust streamers in the core of M86 that connect to
the nucleated dwarf galaxy VCC~882, and suggest a recent interaction
between the pair that may have contributed to the asymmetry of M86's
optical isophotes.

At X-ray energies, M86 has been observed by  {\it Einstein} (Forman \etal\
1979; White \etal\ 1991), {\it Ginga} (Takano
\etal\ 1989), {\it EXOSAT} (Edge 1990), {\it ASCA} (Matsushita \etal\ 
1994), and {\it ROSAT} (B\"{o}hringer \etal\ 1994; Rangarajan \etal\
1995).  
Recently, Finoguenov \etal\ (2004) presented {\it XMM-Newton}
observations of M86 and concluded that its unusual morphology is
due to an interaction with an X-ray filament rather than the Virgo
ICM.  However, their conclusions were based on the large
separation between M86 and M87 along the line of sight of $2.4 \pm 1.4$~Mpc reported
by Neilsen \& Tsvetanov (2000), which is inconsistent with the more
recent result from Mei \etal\ (2007) who find a separation of $0.4\pm
0.8$~Mpc.

We report here on a mosaic of nine {\it Chandra} observations of M86
and the surrounding field, totaling 240~ksec of exposure.
The observations and data reduction techniques are described in
\S~\ref{sec:obs}.  The X-ray image is presented in \S~\ref{sec:img},
and results on temperature and abundance structure from spectral
analysis are given in \S~\ref{sec:spec}.  In \S~\ref{sec:discuss}, we
discuss some of the more interesting features of M86, and place
constraints on its orbit.  In particular, we give new results
on the extent of the stripped tail, and measure the density profile of
an X-ray brightness edge seen to the southeast.  Our results are
summarized in \S~\ref{sec:summary}.

We assume a distance to the Virgo cluster of 16 Mpc throughout,
consistent with the latest results from Mei et al.\ (2007), which
gives a scale of 0.08 kpc/\arcsec\ for $\Omega_0 = 0.3$,
$\Omega_{\Lambda} = 0.7$,
and $H_0 = 70$~km~s$^{-1}$~Mpc$^{-1}$.  All error ranges are 90\% confidence
intervals, unless otherwise stated.

\section{Observations and Data Reduction} \label{sec:obs}

Table~\ref{tab:obs} summarizes the nine {\it Chandra} observations of
M86, which completely cover
the center of the galaxy, as well as the plume and long tail.
All data were reprocessed from the level 1 events files using the latest
calibration files (as of {\sc CIAO3.3}).  CTI and time-dependent
gain corrections were applied 
where applicable. {\sc LC\_CLEAN} was
used to remove background
flares\footnote{\url{http://asc.harvard.edu/contrib/maxim/acisbg/}}.
After periods with obvious flares were removed, the 
mean rate was calculated for some observed quiescent
time interval, and the program was re-run, forcing the global mean to
equal this quiescent rate.
For observations without obvious flares, time bins that were not
within 3$\sigma$ of the mean were discarded.  The final cleaned
exposure times are given in column~(5) of Table~\ref{tab:obs}.

The emission from M86 and the surrounding Virgo cluster fills the
image field of view for each observation.  We therefore used the
standard {\sc CALDB\footnote{\url{http://cxc.harvard.edu/caldb/}}}
blank sky background files appropriate for each observation,
normalized to our observations in the 10-12 keV energy band. We
combined background maps
for each pointing separately, and
generated exposure maps to account for the pointing offsets and the
different CCD responses.  To generate exposure maps, we assumed a MEKAL
model with $kT = 1$~keV, Galactic absorption, and abundance of 30\%
solar at a redshift $z = 0$, which is consistent with typical results from
detailed spectral fits (see \S~\ref{sec:spec}).

\section{The X-ray Image} \label{sec:img}

\begin{figure}
\plotone{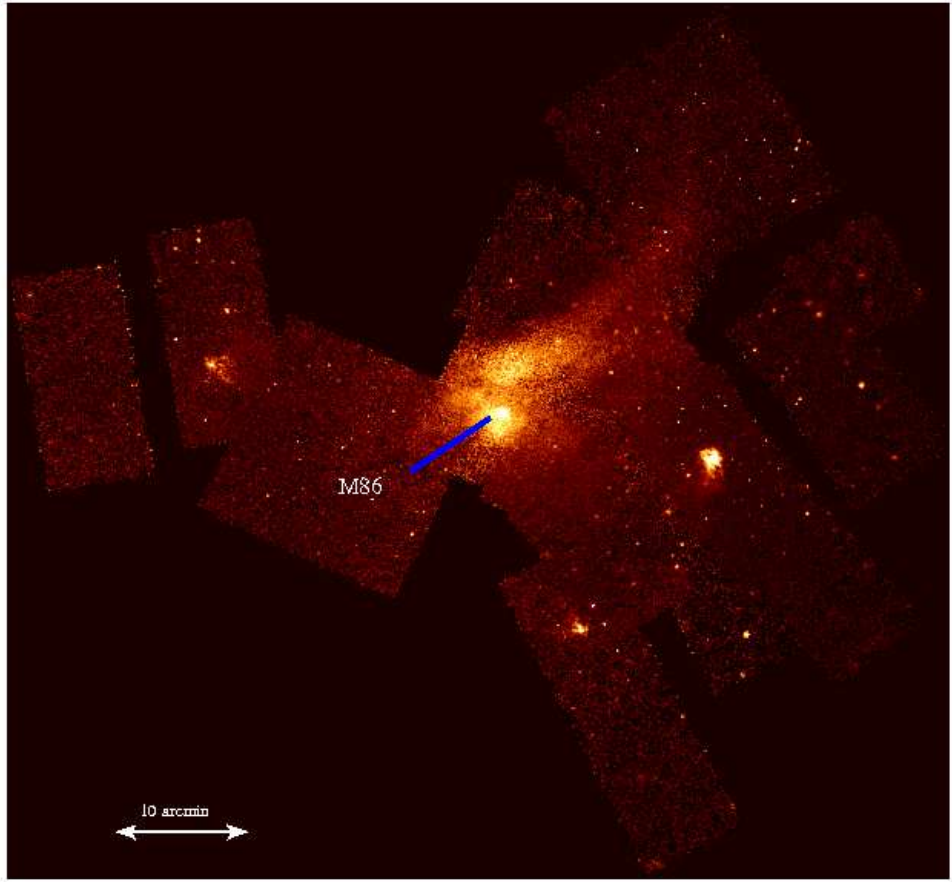}
\caption{Exposure corrected, background subtracted 0.5--2 keV mosaic image of
  the {\it Chandra} observations in the region of M86.  The image has
  been smoothed with a 6\arcsec\ 
  radius gaussian. 
  For each pointing, regions with less than
  10\% of the total exposure for that observation were omitted.
  Complex structure in the diffuse emission is
  observed near and in the stripped tail of M86.  Directly to the
  west of M86, M84 is
  visible, which, aside form showing complicated structure in the
  core, shows a tail of diffuse emission to the south.  Also visible
  are NGC~4388, to the south of M86 and M84, and NGC~4438, directly
  east of M86, both of which show structure in their diffuse emission.
\label{fig:fullimg}}
\end{figure}

The exposure corrected, background
subtracted, smoothed mosaic image is shown in
Figure~\ref{fig:fullimg}.
Several patches of diffuse emission are seen
in M86, as is an impressive tail extending to the northwest.
To enhance the visibility of the diffuse emission we created an image with
bright point sources removed.  
For each {\it Chandra} pointing, regions containing point sources were ``filled in'' using a Poisson distribution whose mean was equal to
that of a local annular background region.
The resulting smoothed image (Figure~\ref{fig:smoimg}) shows:
\begin{itemize}

\item
a long tail extending to the NW, first detected by Forman \etal\
(1979).  The tail is bifurcated at its base, with
the brighter (northern)
extension at a position angle of $305^{\circ}$ (measured from north
to east) and the shorter extension at $285^{\circ}$.  At 18.4\arcmin\
(88.4~kpc) from the center of the bright
emission at the base of the tail (or 21.1\arcmin\ (101.5~kpc) from
the center of M86), the tail turns 
directly north for 6.8\arcmin\ (32.5~kpc).  It then curves to the
west at 26.7\arcmin ~(128.1~kpc) from M86, and finally is no longer detected at
30.6\arcmin\ (150~kpc). There is evidence for a fainter, parallel tail
along the northeastern edge of the tip of the bright tail, discussed below.

\item
a large plume of emission, directly north of M86, at the root of the
long tail (noted previously by Forman et al. 1979).  The
tail emanates from this clump, rather than from M86
itself as one would expect for a continuous stripping process.

\item
a sharp boundary along the northeastern edge of the tail, and an
apparent ``void'' in the diffuse emission just north of the tail.  This
is in stark contrast to the southwestern edge of the tail, where the
surface brightness falls off smoothly.  Rangarajan \etal\ (1995) first
noted this void, and suggested that it is an evacuated Mach cone left
from M86's passage through the outer regions of the Virgo ICM.

\item
an extended halo of X-ray emission associated with M86 with an edge
13\arcmin\ (62.4~kpc) to the southeast.  This edge also is visible in
{\it ROSAT} observations (see Figure~\ref{fig:rosat}).

\item
an asymmetry in the diffuse emission near the core of M86, with an
extension to the south.  Rangarajan \etal\ (1995) suggest that this is
a high pressure region, formed as the leading edge of M86 plows into
the Virgo ICM.

\item
several nearby galaxies.  The elliptical galaxy M84, west of M86,
shows complicated
structure in the diffuse emission in the core related to the radio
outburst, and a tail extending to
the south (Finoguenov \& Jones 2002).  Also
visible are NGC~4388 (south-southwest of M86; Beckmann et al.\ 2004)
and NGC~4438 (east of M86; Machacek et al.\ 2004; Vollmer et al.\
2005). Both also
have diffuse tails, which likely indicate their projected direction of
motion with respect to the Virgo ICM.  Most tails split
into two main streams.  A small patch of faint emission
also can be seen in the region of the disk galaxy NGC~4402 (Crowl et
al.\ 2005).
\end{itemize}

For comparison, the (heavily binned) {\it Chandra} mosaic image is
plotted alongside the {\it ROSAT} and {\it DSS} images in
Figure~\ref{fig:rosat}.  The {\it ROSAT} image clearly shows the
distribution of diffuse emission on larger scales.  In particular, the
contribution from the M87 halo is seen in the southeast, as is
a large halo of diffuse emission around M86 itself, outside the
brightest central regions.  These components are individually
analyzed in our detailed spectral fits (see \S~\ref{sec:dspec}).
Results from these fits indicate that the extended M86 halo is
composed of group gas associated with M86, at a temperature of 
$kT \approx 1.2$~keV.

A close up view of the stripped tail, with scaling and binning chosen
to show its structure more clearly, is presented in Figure~\ref{fig:tail}.
The twisting and curving of the tail at the faint tip may be caused by
turbulence or similar ``weather'' in the Virgo ICM.
The tail is clearly bifurcated
at its base.  Additionally, a fainter
tail, split from the first, is barely visible along the
northern edge of the faint end of the main tail. To test the significance of this
feature, we extracted
the total count rate in evenly spaced bins across the width of the box
shown in Figure~\ref{fig:tail}.  Figure~\ref{fig:tailproj} shows two distinct peaks.  If we take an average count rate from the highest three
bins in each peak and compare it with the average count rate from the
lowest two bins in the gap, we find that the gap is significant at
about 2.5$\sigma$.  
As described below in \S~\ref{sec:displace}, these double streams, as
seen clearly at the base of the tail and possibly at the faint end of
the tail, as well as in other Virgo cluster galaxies in the field, may
be explained as stripping from an inclined aspherical potential.

\section{Spectral Analysis} \label{sec:spec}

In summary, the X-ray images show
 three diffuse emission components: the Virgo ICM (centered on
M87), the extended M86 halo, and the core and tail of M86.  We generate a
temperature map as a guide for detailed spectral fitting to
disentangle the various components.
We assume a galactic absorption of
$N_H = 2.62\times 10^{20}$ cm$^{-2}$ throughout.  Allowing the
absorption to vary did not significantly improve any of the spectral fits.

\subsection{Temperature Map} \label{sec:tmap}

The temperature map was derived using a method employed
by O'Sullivan et al.\ (2005) and Maughan et
al.\ (2006).  For each temperature map pixel
(19.7\arcsec/pix), we extracted a spectrum from a circular region
containing 1000 net counts
(after subtracting the blank sky
background and a component to account for emission from M87, see
\S~\ref{sec:dspec}), up to a maximum radius of 3.3\arcmin\ (on the order of the chip size).
This large maximum
radius was used to obtain sufficient source counts to extend the
temperature map to faint regions of the tail.  
The resulting spectrum was
fit in the 0.6 -- 2.0~keV range with an absorbed APEC model using {\sc
  XSPEC}.  
Data from back-side- (BI) and front-side-illuminated (FI) CCDs 
were treated as separate data groups for each
observation.
For each fit, the abundance was fixed at 26\% solar (consistent with
most of the detailed  
spectral fits; see Table~\ref{tab:spectra}).  The
resulting temperature map is shown in Figure~\ref{fig:tmap} (middle).  Regions
that did not have at least 1000 source counts within the maximum
3.3\arcmin\ radius were excluded from the map, as were regions
in the southeast that were dominated by Virgo cluster emission.
For comparison, we also show the tessellated temperature map in
Figure~\ref{fig:tmap}.  Each bin was fit using only counts from that
area, so the extraction regions are well-defined.  The bins were
generated using the algorithm provided by Diehl \& Statler (2006),
which is a generalization of Cappellari \& Copin's (2003) Voronoi
binning algorithm, and requiring roughly 1100 net counts per bin.
Each bin was fit with a single APEC model, without a component to
model the Virgo ICM background, in contrast to the ``smoothed''
temperature map.  The size of the bins roughly indicate the size of
the extraction regions for the smoothed temperature map pixels in the
same area.  The tessellated and smoothed temperature maps are in good
agreement, except in the regions far from M86 where the contribution
from the Virgo cluster is more important.

In these temperature maps, the remarkably sharp
boundary between the cool galaxy gas and that of the cluster on the
northern edge of the tail is apparent (Figure~\ref{fig:tmap}).  At least four
distinct cold clumps can be seen: one centered
on M86, one 1.41\arcmin\ (6.8~kpc) south
of M86, one
elongated clump 2.45\arcmin\ (11.7~kpc) east of M86, and a larger cool
plume 3.63\arcmin\ (17.4~kpc) to the north. 
Each clump corresponds to a bright peak in the X-ray image.
Additionally, the cooler tail extends to the northwest.
To the southeast, in the region of the brightness edge discussed in
\S~\ref{sec:halo}, the gas is somewhat hotter ($\sim0.25$~keV) in the
southern half of this feature than in the
northeastern half at the same distance from M86.  This temperature difference
is detected in spectra from non-overlapping regions in these two
areas, indicating that the difference is not an artifact of the
technique used to generate the temperature map. More detailed fits are
discussed in \S~\ref{sec:dspec}.

\subsection{Detailed Spectra} \label{sec:dspec}

Based on the derived temperature map and the X-ray image, we
defined 23 regions for detailed spectral analysis (see
Figure~\ref{fig:regmap}).
 To include data from a particular observation in the spectral
 analysis, we required that it contribute at least 200 counts (each
 data group was binned such that there were
at least 40 counts per bin).
A summary of the regions and spectral model for each
region are given in Table~\ref{tab:spectra}.
R1~-~R4 are centered on the individual clumps of emission near
M86, while R5~-R10 trace the long tail 
NW of M86 (hereafter regions will be labeled as 'R\#', with the
number defined by column~1 of Table~\ref{tab:spectra}). 
R11 roughly corresponds to the ``shocked region'' indicated by
Finoguenov et al.\ (2004; see their figure~7).  There are several
regions to either side of the M86 tail.  R17, R18, R19,
\& R23 measure the large scale diffuse components from
M86 and M87, outside the central region where the cool gas is observed (note
that R23 is a combination of five disjoint regions, see Figure~\ref{fig:regmap}).
Finally, R14~-~R16 contain M84 and its tail of diffuse emission.
Each region was initially fit with an
absorbed APEC model over the 0.6-5.0 keV range, with an additional
APEC component to model emission from the Virgo ICM as in \S~\ref{sec:tmap},
with the abundance allowed to vary.
Additional components were added, as needed, to improve the fits (see
Table~\ref{tab:spectra}.  For
example, a single temperature model gave large residuals at high
energies when fit to the regions corresponding to the cores
of M86 and M84 (R1 and R14, respectively).  Since each of these regions is expected to have a
contribution from unresolved X-ray point sources, a power law component
was added to the model, which greatly improved the fits
(and gave a photon index consistent with $\Gamma \approx 1.5$, as
expected, see e.g. Sarazin et al.\ 2003). 
For regions where this power law component could not
be well-constrained, we fixed the photon index at $\Gamma = 1.5$.
Nine regions
showed an improved fit when another APEC model was added (in such
cases the abundances of each component were constrained to vary together).  
  Data from each pointing were treated as 
separate data groups, with normalizations for each pointing allowed to
vary independently.   As an example, the two
temperature plus power law fit to the 
spectra for the core of M86 (R1) is shown in Figure~\ref{fig:m86spec}.

Each of the three main
components of the diffuse emission (the Virgo ICM, M86 group gas,
and the cooler central M86 gas) was
detectable in some subset of the regions shown in
Figure~\ref{fig:regmap}. In R13 it was possible to
fit all three components simultaneously.  The Virgo and M86 halos cover
most, if not all of the FOV.  It was therefore necessary to include
these components when doing detailed spectral fits, even in regions
where they could not be modeled directly due to inadequate statistics.
To account for the varying Virgo contribution, we took the
normalizations of the best fit APEC models 
to R18 and R23 and linearly interpolated across the field
based on the distance from the center of M87. 
Although this only gives a rough estimate, we note that it does not
affect our results significantly.  For regions closer to M87 (e.g.,
R19) the interpolation gives a value very close to that measured
from R18, which is an appropriate local background.  Similarly,
regions near the faint end of the M86 tail (e.g., R10)
have normalizations close to that of R23, which again, is an
appropriate local background for this area.  For most regions of
interest,
M86 is bright enough that the contribution from Virgo is negligible, and
the accuracy of interpolating is therefore sufficient.  

A similar approach was used to
model the contribution from the halo around M86.  Although, in
general, the normalization of this component decreased with distance
from the core of M86, the variation across the field was more
complicated than for Virgo, which is expected since this gas is in the
process of being stripped from M86.  We included
a fixed component, with $kT = 1.2$~keV, abundance set equal to that of
the main component, and variable normalization, in
regions where the M86 halo temperature could not be determined
directly (see column~2 of Table~\ref{tab:spectra}). The effects of
accounting for this component can be seen, for example, by comparing
lines~1~\&~2 in Table~\ref{tab:spectra}.  Including the hotter 1.2~keV
component results in a lower best-fit temperature for the cooler gas,
and gives a better statistical fit to the data.  In general, our
results are consistent with Finoguenov \etal\
(2004), who derive X-ray temperatures and abundances for the central
region, including M86 and the plume from {\it
  XMM-Newton} observations.

We note that for the cores of M86 and M84 (R1 and R14 in
Table~\ref{tab:spectra}) the abundances are poorly
determined, despite the large number of net counts.  
Using a VAPEC model to allow for non-solar abundance ratios did not
improve the fits.
The failure of these models may be due to
abundance variations within the defined regions.  For instance, the
abundance map given by Finoguenov et al.\ (2004) shows significant
variation near the core of M86, even on relatively small scales.
Additionally, the models do not allow for multiphase gas, although the
temperature most likely varies significantly in the cores.

Using the detailed spectra, we can derive the approximate mass in 
cool gas for the base of the tail (the ``plume'') and the core of
M86.  Using the temperature and abundance measured in R4, and
assuming a prolate spheroid geometry with major and minor axis lengths of
12.3~kpc and 7.9~kpc, we find a density of $n_{\rm plume} \approx 8.2
\times 10^{-3}$~cm$^{-3}$ and a mass $M_{\rm plume} \approx 7.4 \times
       10^8 \, {\rm M_\odot}$.  Similarly, in the core of M86, we find
$n_{\rm core} \approx 6.2 \times 10^{-3}$~cm$^{-3}$ and $M_{\rm core}
       \approx 7.4 \times 10^8 \, {\rm M_\odot}$ within a sphere of
       radius 9.6~kpc.  
       Assuming a
       cylindrical geometry for the tail, and considering only the cool
       gas components in R5 - R9, we find a typical
       density 
       of $n_{\rm tail} \approx 1.5 \times 10^{-3}$~cm$^{-3}$ and a
       total gas mass of $M_{\rm tail} \approx 1.7 \times
       10^9 \, {\rm M_\odot}$.  
       The mass in stripped gas is therefore more than three times
       the gas mass in the core, which supports the scenario where 
       a significant fraction of the M86 ISM is removed rapidly due to
       ram pressure
       stripping (Takeda \etal\ 1984).  Rangarajan \etal\ (1995) find similar results for the plume
       and the core.
       However, we find consistent abundance values in the core and in
       the plume within the errors, while Rangarajan \etal\ (1995)
       find a higher abundance in the plume.
       This disagreement is likely due to
       the multiphase model we use, which includes a separate
       component for the 1.2~keV M86 halo emission.

\section{Discussion} \label{sec:discuss}

\subsection{The Ram Pressure Stripped Tail of M86} \label{sec:tail}

\begin{figure}
\plotone{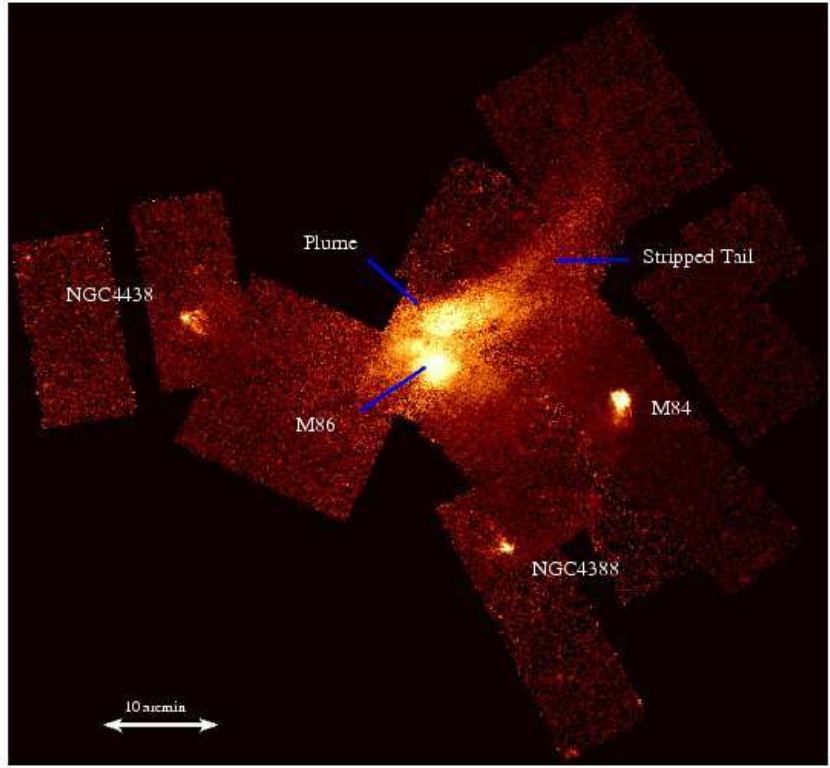}
\caption{Same as Figure~\ref{fig:fullimg}, but with different scaling
  and a larger smoothing kernel (9\arcsec) to highlight the structure
  at the low surface brightness end of the M86 tail and to 
  show the faint M84 tail more clearly.  Point sources
  have been removed (see text for
  details).  The main M86 tail, after turning
  from northwest to north, appears to turn back to the west near the very tip.
\label{fig:smoimg}}
\end{figure}

The most striking feature in Figure~\ref{fig:smoimg} is the long tail
of emission extending to
the NW from M86.  M86 has a line-of-sight velocity of -244$\pm
5$~km~s$^{-1}$, while M87's
is 1307$\pm 7$~km~s$^{-1}$ (Smith et al.\ 2000).  
Therefore, M86 is traversing the Virgo cluster at 
$v_{\rm M86} > 1550$~km~s$^{-1}$ (about Mach 2 for $kT = 3$~keV).
The tail naturally forms due to ram pressure stripping of the M86
corona by the Virgo
cluster ICM.  
Finoguenov \etal\ (2004) suggested that the tail formed due to
interactions with a filament rather than the Virgo ICM, though their
result was based on 
older distance measurements that placed M86 outside the virial radius
of the Virgo cluster.
If we assume that M86 is bound to
the Virgo cluster, its large line-of-sight relative velocity allows us
to constrain the separation between M86 
and M87. 
Using the M87 mass profile detailed in \S~\ref{sec:orbit},
we find that a free-fall
velocity of 1500~km~s$^{-1}$ corresponds to a separation from M87 of 0.5~Mpc
(the projected separation is 0.35~Mpc).
This is an upper-limit, since the total
relative velocity must be at least as large as the line-of-sight
relative velocity.  
This separation is consistent with recent surface brightness
fluctuation distances of Virgo cluster galaxies,
which give a distance between M86 and M87 of $0.4\pm 0.8$~Mpc (Mei et al.\
2007).

We also can place a lower-limit on the
length of the long stripped tail using the mass profile of M87 and the
line-of-sight velocity.  The maximum free-fall velocity of M86 from infinity
at the projected separation of 0.35~Mpc
is about 1680~km~s$^{-1}$.  Using this as an upper-limit on its current
3D velocity, we find that the angle between the direction of motion of
M86 and the line-of-sight is $\theta \la 23^{\circ}$.  Assuming that the
stripped tail is aligned with M86's current direction of motion, and
given that the
length of the tail in the plane of the sky is 150~kpc
(0.51$^{\circ}$), we find a 
lower-limit on the actual length of the tail of $L_{\rm tail} \ga
380~$kpc, making this the longest ram pressure stripped tail
presently known.


The long stripped tail originates in the plume of diffuse
emission located directly north of M86.  For a continuous stripping
process, one would expect the tail to extend from M86 itself.  This
separation between M86 and the plume can be explained by
rapid recent stripping, in which a significant fraction of the
remaining gas in M86 is
rapidly stripped when the ram pressure stripping condition is met.  A
detailed discussion of this process is given in \S~\ref{sec:displace}

As seen from Figure~\ref{fig:regmap} and Table~\ref{tab:spectra}, the
stripped tail shows a general trend of cooler gas in and near the
plume (with temperatures in the 0.8--0.85~keV range) and warmer gas at
the tip (in the 0.9--1.2~keV range).  The temperature structure of the
tail is consistent with a ram pressure stripping model, where, as
M86 falls into the Virgo cluster, the hotter, higher entropy group gas is
stripped first due to interactions with the Virgo ICM, followed by the
cooler, lower entropy M86 ISM, which is removed rapidly once the
stripping condition is met.

\subsection{The Extended M86 Halo} \label{sec:halo}

\begin{figure}
\plotone{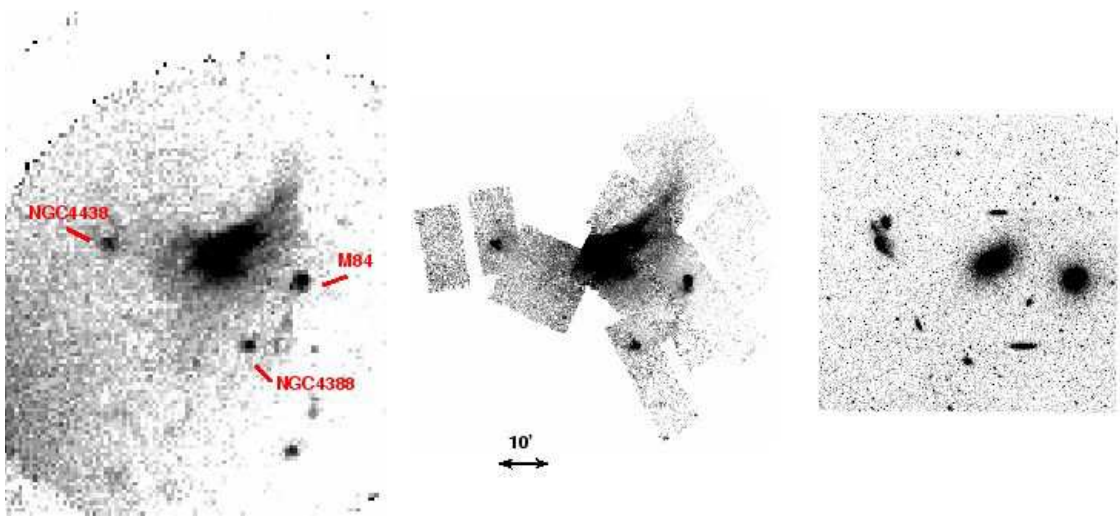}
\caption{
  (left) {\it ROSAT} PSPC view of M86.  Diffuse emission from M87/Virgo is
  clearly visible in the southeast, as is an extended halo of diffuse
  emission associated with M86.  Also shown are the {\it Chandra}
  (center) and {\it DSS} (right) images of the same region.
\label{fig:rosat}}
\end{figure}

\begin{figure}
\plotone{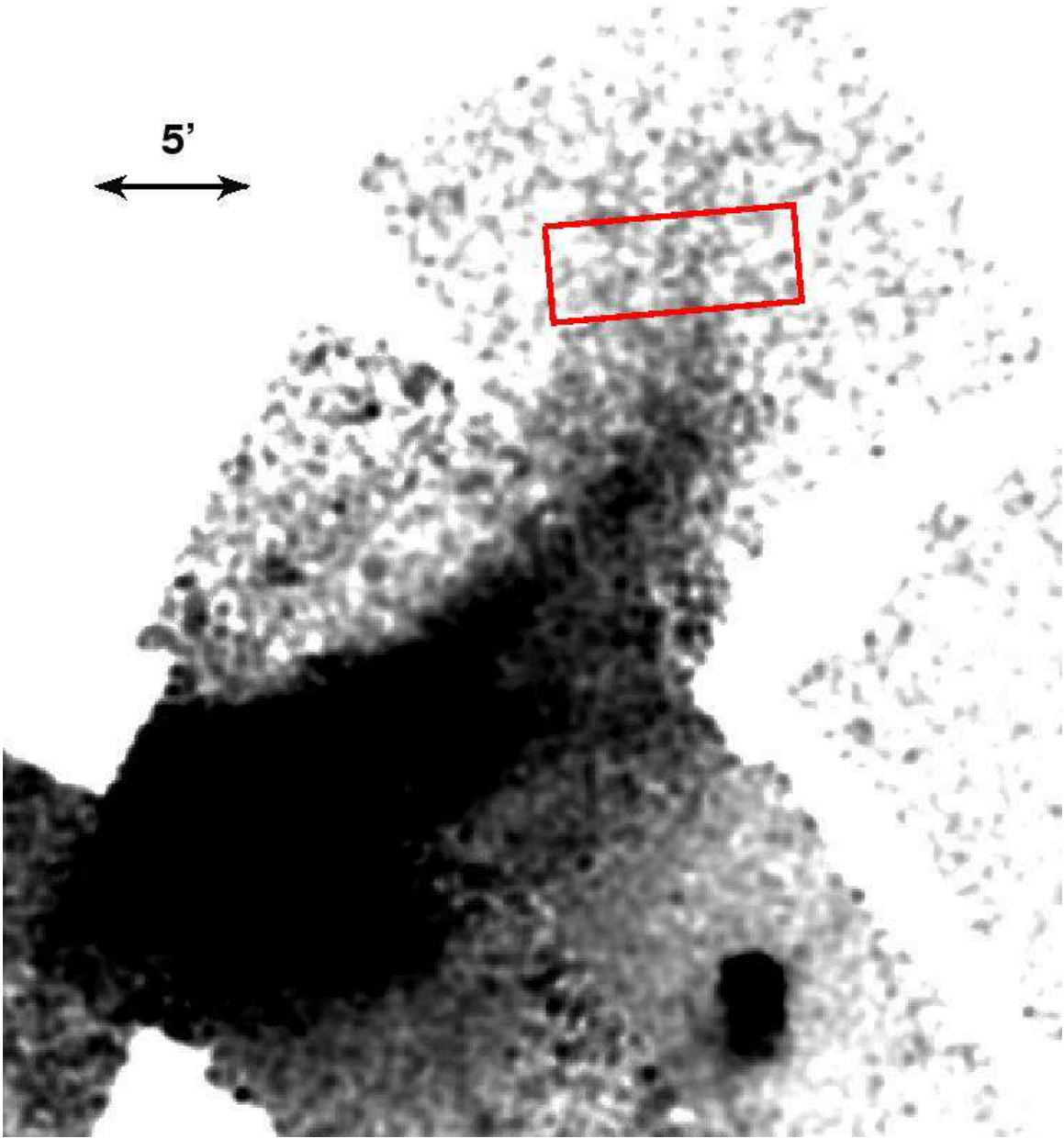}
\caption{
Close up view of the long stripped tail, in the 0.5--2~keV band and
smoothed with
a 16\arcsec\ radius gaussian.  The tail clearly turns from
a NW direction to the north, and back directly east at the very tip.
Furthermore, there appears to be a faint secondary tail along the
northern edge of the faint end of the main tail, which traces its
path.  The red box indicates the region used to generate the plot in
Figure~\ref{fig:tailproj}.
\label{fig:tail}}
\end{figure}

\begin{figure}
\plotone{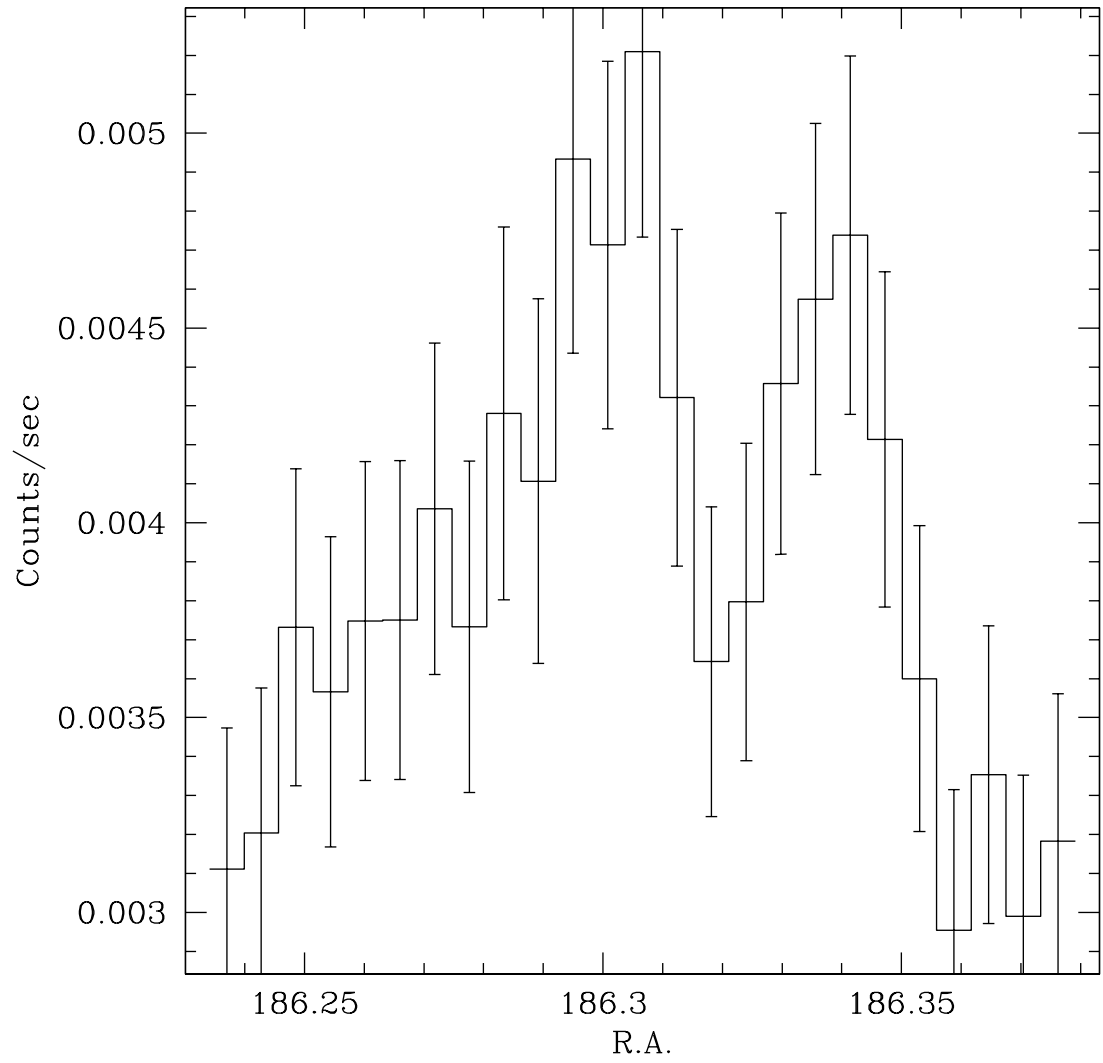}
\caption{Plot of the total count rate in evenly-spaced binned regions
  taken across the width of the box shown in Figure~\ref{fig:tail}.
  The x-axis gives the R.A. of each bin on the southern edge of the box.
\label{fig:tailproj}}
\end{figure}

\begin{figure}
\plotthree{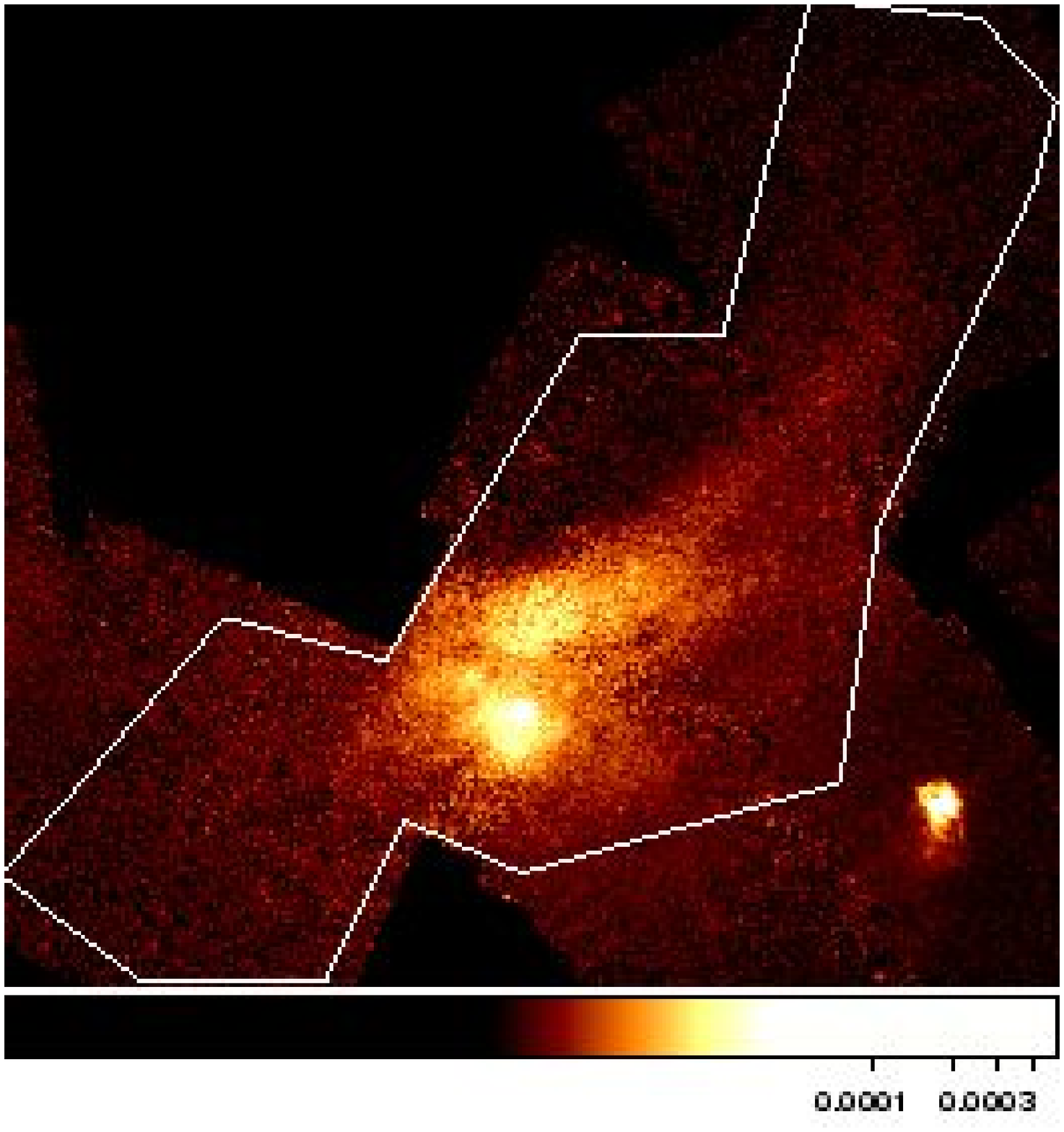}{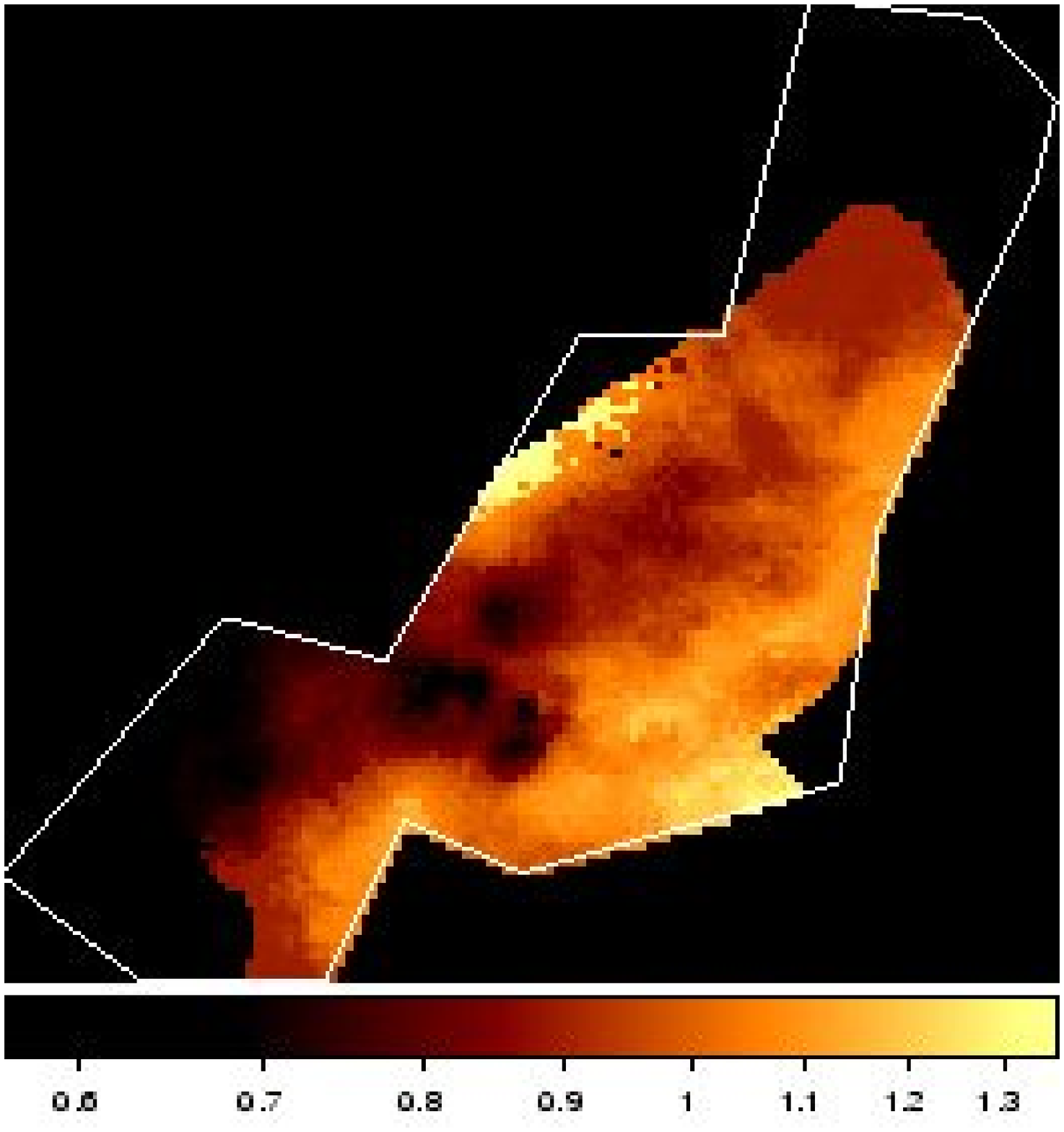}{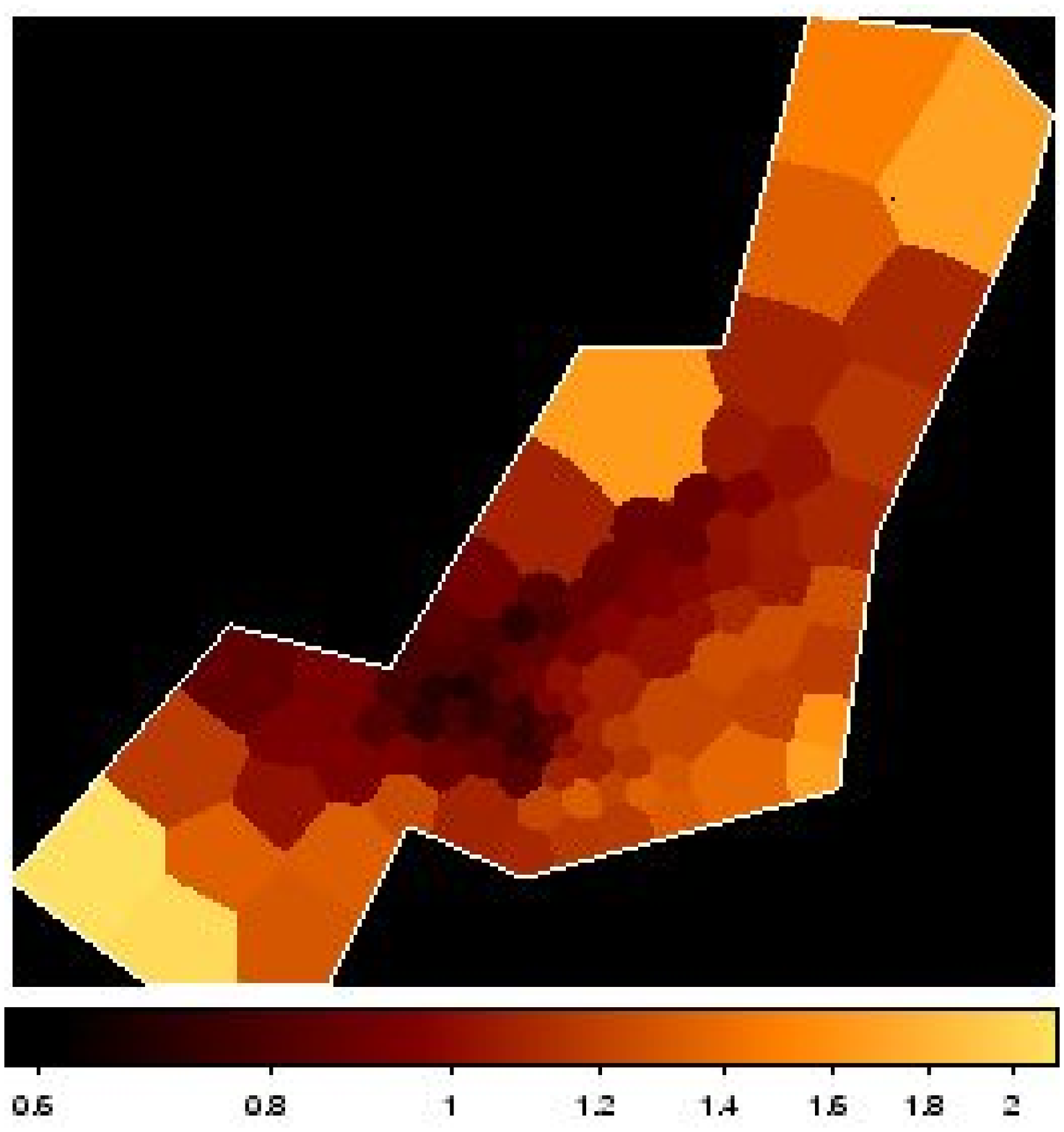}
\caption{
{\it Left Panel:} A closer view of the central region of
Figure~\ref{fig:smoimg}, with scaling chosen to highlight structure in
the central region and nearby diffuse emission.  The region used to
extract the temperature map is overlain. {\it Center Panel:} Derived
temperature map of the same region.  The contribution from the Virgo
ICM has been taken out.
Zero-valued pixels indicate
areas that required regions larger than the selected maximum area in
order to
include the minimum net counts (see \S~\ref{sec:tmap}).
Pixels in faint regions which gave an anomalous fitted temperature
were also removed.  The colorbar under the right panel gives the
temperature in keV. {\it Right Panel:} Tessellated temperature map.
Each bin contains roughly 1100 net counts.  The counts in each bin
were fit with a single APEC model.  The contribution from the Virgo
ICM has {\it not} been taken out, leading to higher temperatures farther
from M86 as compared to the center panel.  The size of the bins
roughly indicate the size of the extraction region for each pixel in
the center panel.
\label{fig:tmap}}
\end{figure}

\begin{figure}
\plotone{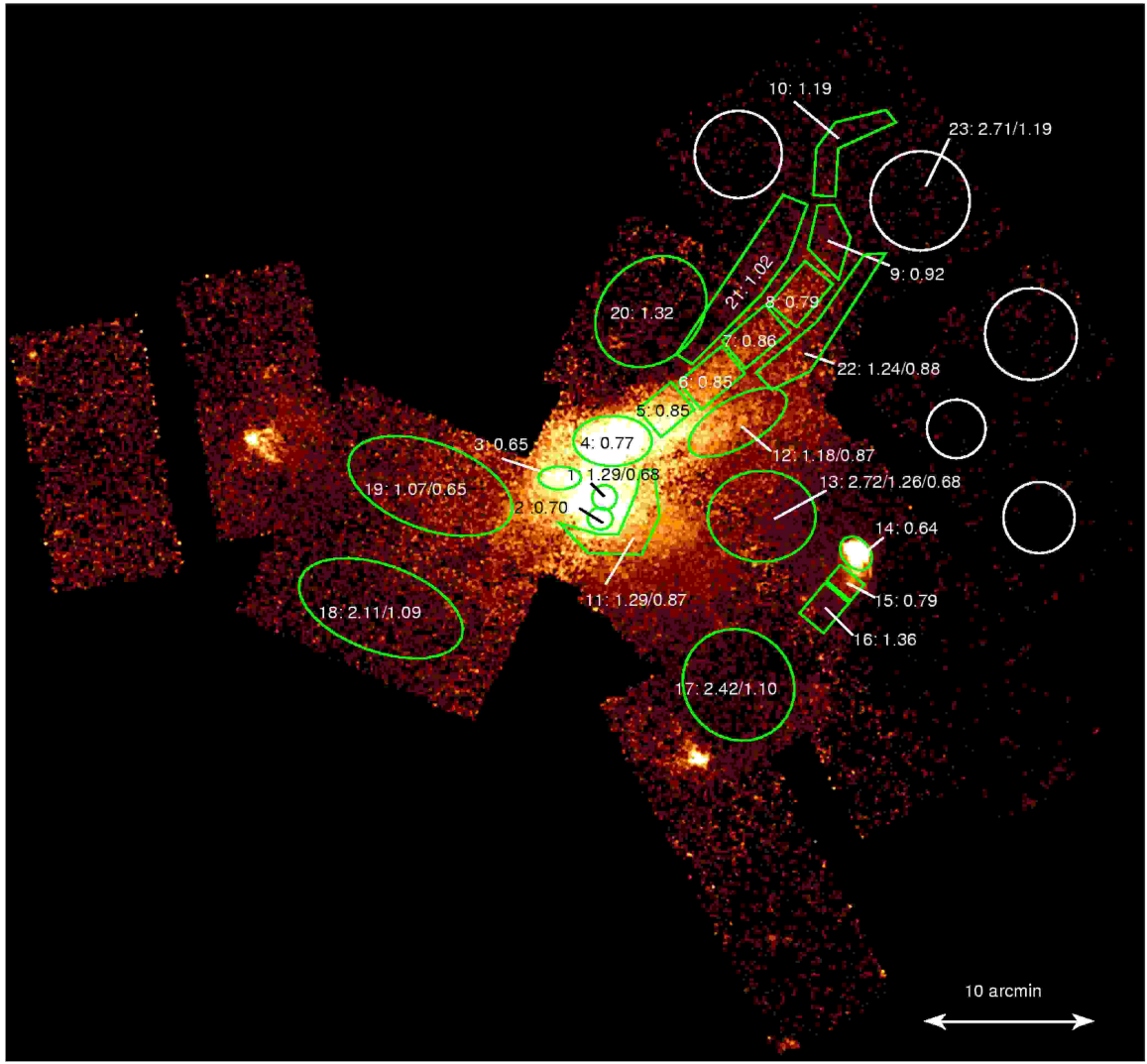}
\caption{
  Same as Figure~\ref{fig:smoimg}, but with the regions used for
  extracting spectra overlain.
  Single regions for spectral fitting are
  shown in green, while the regions shown in white are fit as one combined
  region.  Table~\ref{tab:spectra} gives the region number and the
  best-fit temperature(s).
\label{fig:regmap}}
\end{figure}

\begin{figure}
\includegraphics[angle=270,width=\linewidth]{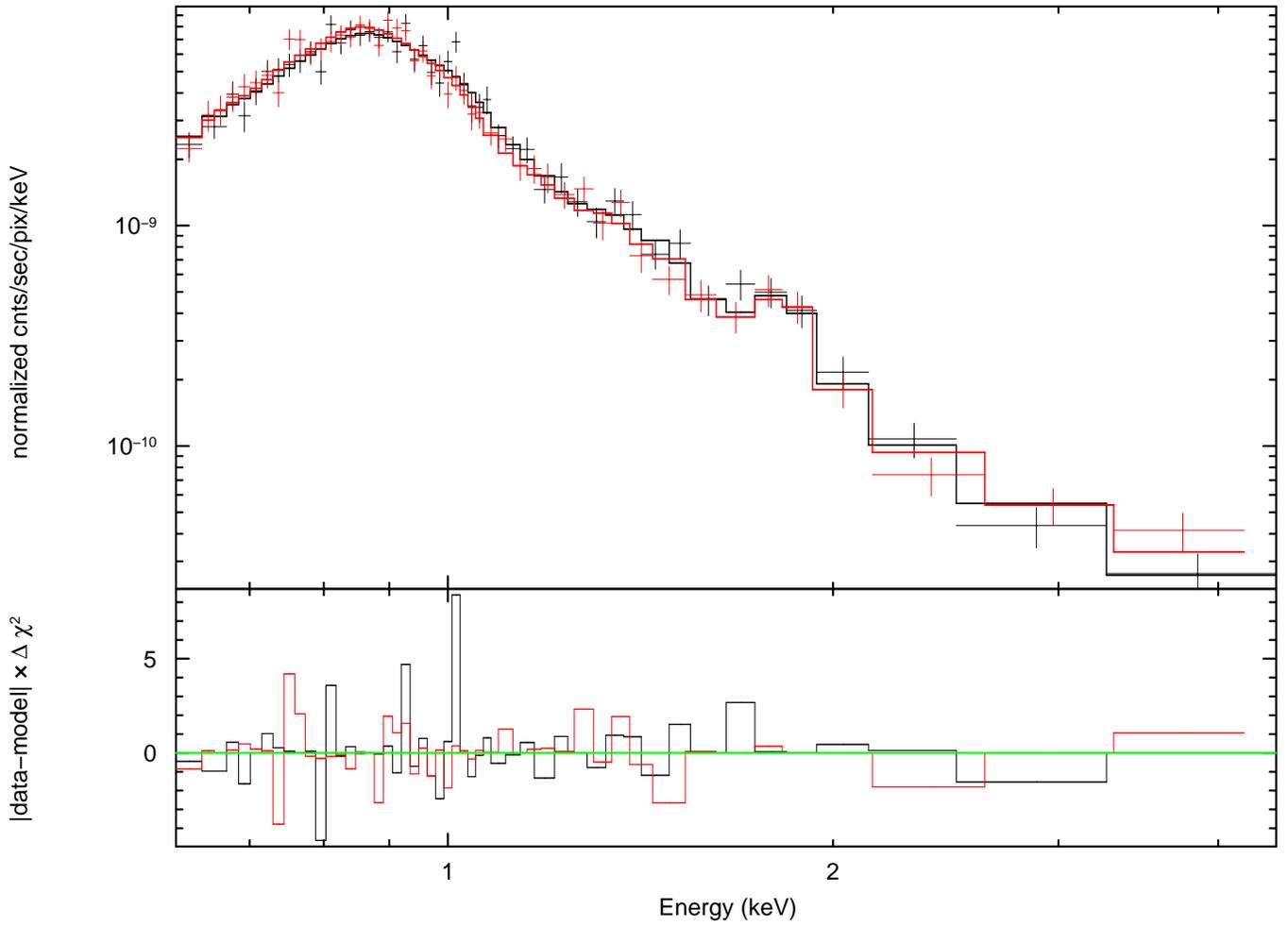}
\caption{
Data (crosses) and fitted model (histogram) for the core of M86,
R1 in Table~\ref{tab:spectra}.  The model is a two-temperature thermal
plasma with fitted temperatures and abundance (given in the second row
of Table~\ref{tab:spectra}), plus an unresolved
power law
component with photon index of 1.5 and a fixed thermal plasma
component to model emission from the Virgo ICM, all with Galactic absorption.
The data are from two S3
observations, OBS-IDs 318 and 963 (see Table~\ref{tab:obs}).
\label{fig:m86spec}}
\end{figure}

\begin{figure}
\includegraphics[angle=270,width=\linewidth]{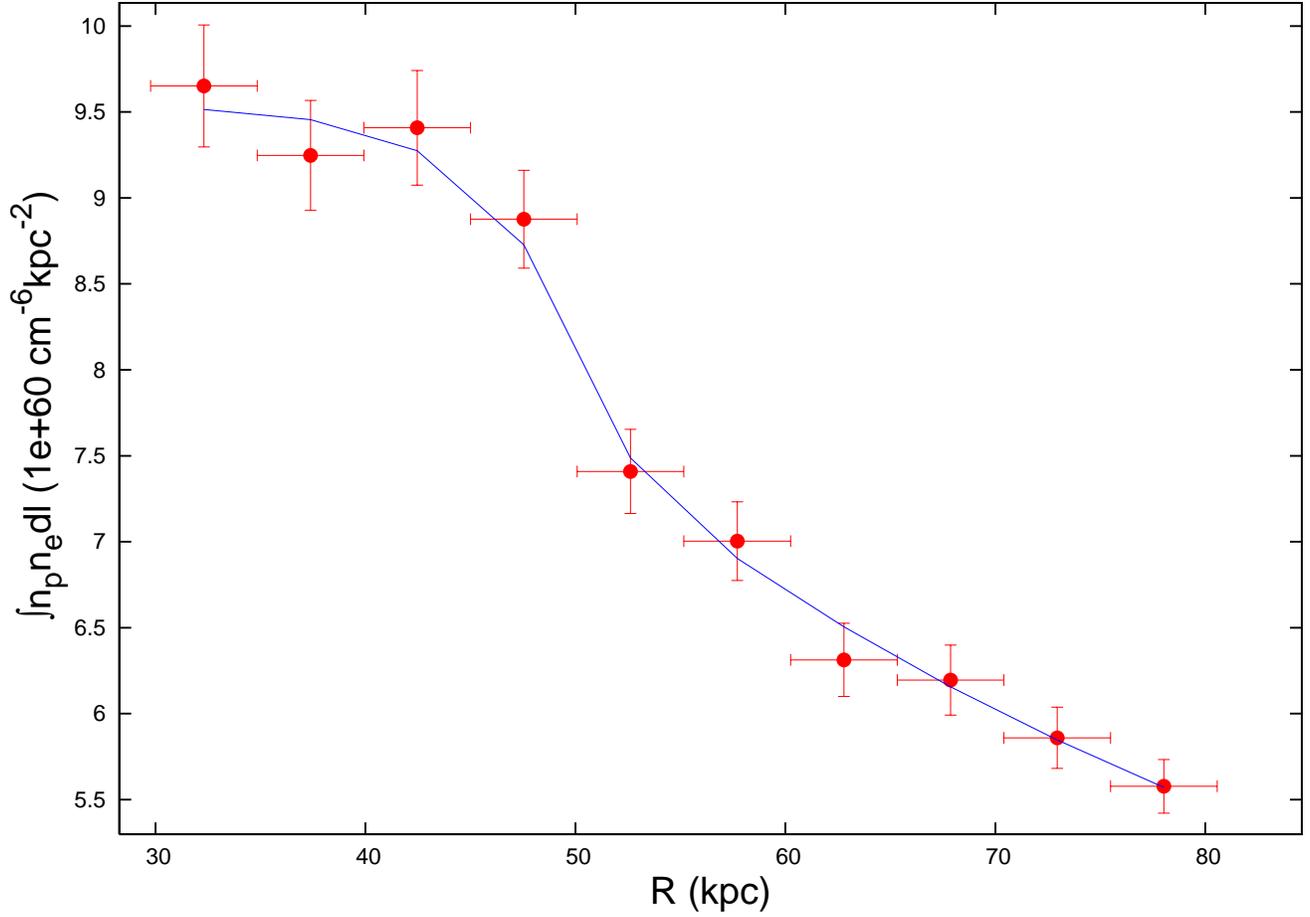}
\caption{
Emission measure profile of the northern half of the bright edge seen in the extended
group emission southeast of M86 (shown most clearly in the {\it ROSAT} image
in the left panel of Figure~\ref{fig:rosat}). The x-axis gives
the radius from the apparent center of curvature defined by the
feature.  The best fit two power law density jump model is given by the
solid line.
\label{fig:nedge}}
\end{figure}

The {\it Chandra} and {\it ROSAT} X-ray images
(Figure~\ref{fig:rosat}) show an 
extended halo associated with M86, with a sharp
brightness edge to the southeast in the direction of
M87.  
Since M86 is traversing the Virgo cluster supersonically (see
\S~\ref{sec:tail}) we expect a shock to be driven in the Virgo ICM,
possibly producing a brightness edge similar to that seen in the
southeast.
We attempted to measure the density jump associated with the shock in
the following way.
We extracted the {\it Chandra} 0.5--2.0 keV
surface brightness profile of this edge in two roughly equal sectors
since the morphology of this feature is irregular in the south
(see Figure~\ref{fig:rosat}), and since the gas temperature differs 
in these two regions (see \S~\ref{sec:tmap}).  The emission
measure profile for 
 the northern sector is shown in Figure~\ref{fig:nedge}.  Distance is
 measured from the center of curvature of the apparent edge, which is
 10~kpc (2.16\arcmin) east of M86.  This profile was fit with a
 spherical gas density model consisting of two power laws.  The free
 parameters were the normalization, the inner ($\alpha$) and outer
 ($\beta$) slopes, the 
 position of the density discontinuity ($r_{\rm break}$), and the
 amplitude of the jump ($A$).  We assumed that the gas is isothermal with
 $kT = 1$~keV and that the abundance is constant at 30\% solar,
 consistent with results from spectral fits (see \S~\ref{sec:spec}). 
 For the best fit model (see Figure~\ref{fig:nedge}), we find $\alpha =
 0.49^{+1.14}_{-0.71}$, $\beta = -0.81^{+0.14}_{-0.12}$, $r_{\rm
   break} = 52^{+5}_{-6}$~kpc, and
 $A = 1.3^{+0.3}_{-0.4}$.  Similar results are found from the southern
 sector, but
 with larger errors.  The lack of a well-defined edge is consistent
 with our findings for the orbit of M86 (see \S~\ref{sec:orbit}).
 In particular, for all of the likely orbits, M86 is moving to the southeast,
 such that our 
 lines of sight pass through the Mach cone. To see a sharp edge, our
 line of sight must be tangent to the shock front.

\subsection{Constraints on the Orbit of M86} \label{sec:orbit}

Knowing the orbit of M86 is the key to understanding its interaction
with the Virgo cluster.  The ram pressure stripped tail reveals the
motion of the galaxy on the sky.
The length and direction of the tail, together with the large
line-of-sight speed of M86, constrain its orbit.  The large luminosity
of M86 suggests that it dominates the associated in-falling subgroup
(Schindler \etal\ 1999).  In the following, we assume M86 is
bound to the Virgo cluster. For the purpose of calculation, the
gravitational potential of the
Virgo cluster is treated as a spherical NFW potential (Navarro, Frenk
\& White 1997), with a virial radius of 1.3 Mpc (Evrard, Metzler \&
Navarro 1996) and a concentration parameter of 4.5 (Neto et al.\
2007), appropriate for a $\sim 3$ keV cluster.
The gravitating mass was normalized to match the total mass within 320
kpc of $4.4\times10^{13}\ {\rm M_\odot}$ (Schindler \etal\ 
1999; scaled to a distance of 16 Mpc).  For this potential, the escape
speed from a radius of 351 kpc (the projected separation of M86 from
M87) is 1677 $\rm km\ s^{-1}$, not much larger than M86's line-of-sight
speed of 1550 $\rm km\ s^{-1}$ relative to M87.  The empirical mass
distribution of Schindler (1999) gives the same escape speed at 351~kpc
as our model potential.  We note that the gravitational potential of
the Virgo cluster is not well 
constrained at distances from the cluster center comparable to or
greater than its virial radius.  Furthermore, the truncated NFW
potential is highly simplified, ignoring mass beyond the virial
radius, departures from spherical symmetry, and the dynamic state of
the cluster due to continuing in-fall.  Since some of the limits we derive here
are sensitive to the poorly known potential at large radii, they
should be treated as indicative rather than quantitative (outside the
context of the model).
The material in this section is supplemented
by a more detailed discussion in Appendix~\ref{ap:orbit}.

We consider radial orbits first, as suggested by previous studies (e.g.,
Forman \etal\ 1979; White \etal\ 1991).  
In order for the gas tail to point away
from the cluster center, the galaxy must be inbound.
It can easily be shown that the line-of sight velocity, $\vlos$, is
maximized for some radius $r$ greater than the observed separation $s$
(see Appendix~\ref{ap:orbit}).
For a marginally bound (zero energy) radial orbit in the potential
described above, the maximum value of $|\vlos|$ occurs when M86 is
close to twice its projected distance from the cluster center, giving
$|\vlos| = 1205\ \kms$, less than the observed value of $1550\ \kms$.
Thus, unless M86 is significantly unbound 
from the Virgo cluster (or the potential is incorrect), its
line-of-sight speed is inconsistent with radial and nearly radial orbits.

It is convenient to specify more general orbits in terms of their
inner and outer turning radii, $\ri$ and $\ro$, respectively.  
We constrain the possible orbits of M86 by placing limits on these
parameters.  Coarse limits can be placed on these radii by considering
energy arguments alone.
Assuming that M86 is bound to the Virgo cluster, we
find $\ri \la 489$~kpc and $\ro \ga
3.5$~Mpc, or $\simeq 2.7$ times the virial radius of the Virgo cluster
(see Appendix~\ref{ap:orbit}).  These limits can be further restricted
by considering the range of possible viewing directions for each point
on an orbit.  There can be zero, two, or four possible
viewing directions that would place M86 at the observed separation
from the cluster and give it the observed line-of-sight velocity (see
Appendix~\ref{ap:orbit}). Figure~\ref{fig:mb} shows a range of marginally
bound orbits ($\ro = \infty$), with regions that meet these conditions
marked in color.  The range of an orbit where these conditions are
met shrinks as $\ro$ decreases, \ie, as the orbit becomes more
tightly bound.  For the marginally bound orbits, the full range of
$\ri$ for which $\vlos$ can attain its observed value is $247 < \ri < 395$
kpc (the range represented in Figure~\ref{fig:mb}).  For more tightly
bound orbits, the acceptable range of $\ri$ is reduced.

The lower limit on $\ro$ is increased if we consider the extent of the
ram pressure stripped tail.  The projected orbit must extend to at
least the distance ($\sim 100$~kpc) that the tail projects beyond M86
in the direction away from the cluster center. Locations on the orbits where
the line-of-sight speed can attain $-1550\ \kms$ and these conditions
also are met are shown in red in Figure~\ref{fig:mb}.  We see that the
range of possible locations for M86 on these orbits is tightly constrained.
We repeated this analysis for a number of values for the
outer turning radius.  The range of potential orbits and locations for
M86 shrinks with decreasing $\ro$ and no orbits were found to meet
these criteria for $\ro \lesssim 8.2$ Mpc (6.3 virial radii).

Figure~\ref{fig:projmb} shows projections onto the sky of the marginally
bound orbits (corresponding to the midpoints of the red regions of
Figure~\ref{fig:mb}) overlaid on the {\it Chandra} 0.5 -- 2
keV image, with M86 at the representative location for each orbit.
It is evident that the possible orbits for M86 can be pruned further.
For example, the orbit on the lower right in Figure~\ref{fig:projmb}
does not reach far enough north to produce the remote part of the gas
tail.  At the other limit, the position angle of M87 measured from
M86 is $116^\circ$ (east from north), generally eastward of these
orbits.  Overdense stripped gas tends to fall towards the cluster
center, in that direction, so that some stripped gas can reasonably
lie to the east of the orbit.  Stripped
gas cannot lie west of the orbit.  Thus the orbits that cross the
region to the east of the tail and north of M86, where there is
no sign of stripped gas, are unlikely candidates for the orbit of
M86.

Note that the orbits in Figure~\ref{fig:projmb} are not simply
related to their inner turning radii.  The orbit corresponding to the
smallest value of the inner turning radius appears second from the
right at the top of the figure.  At first, orbits for increasing
values of $\ri$ lie to the left of this, but at a value of $\ri$
approaching (but less than) $s$, the acceptable viewing direction
flips from outward to inward (\ie, the acceptable sign of $a'$
changes, see Appendix~\ref{ap:orbit}).  At this point, the projected 
position of the orbit shifts 
from leftmost to rightmost at the top of Figure~\ref{fig:projmb}.
Whereas the apparent location where the orbit passes through the
virial radius was moving outward, it now moves inward as $\ri$
increases, coming into the field of view for the largest values of
$\ri$ here.  This results in disjoint ranges of possible orbits for
M86.  Thus the marginally
bound orbits with inner turning radii close to $\ri \simeq 260$ kpc and
those with inner turning radii in the range $330 \la \ri \la 380$~kpc
are best suited to model the orbit of M86.

Applying the same criteria to the more tightly bound orbits yields
the collection of possible orbits shown in Figure~\ref{fig:m86orb}.  We
have not attempted to be highly selective or exhaustive, since that
requires a well defined model for the formation of the gas trail.  For
values of the outer turning radius in the lower end of the acceptable
range, the shrinking range of possible orbits and locations excludes
orbits with inner turning radii $\ri > s$.  All of the acceptable
orbits for M86 are weakly bound to the Virgo cluster, with the most
tightly bound having outer turning radii at $\simeq 8.8$ Mpc.  All of
the possible locations for M86 lie only a little farther from M87 than
M86 does in projection.  All are close to the plane of the sky,
ranging from 167 kpc closer than M87 to 263 kpc farther than M87 from
the Sun at the extremes.  M86 must also be close to the pericenter of
its orbit.  The direction of motion of M86 is close to our line of
sight, within $16^\circ$ -- $23^\circ$ of it, for all of the cases
illustrated.  

M86 is traversing the Virgo cluster supersonically ($\vlos =
-1550$~\kms\ alone is
almost twice the sound speed $\sound$, which is taken to be 850 \kms).
Therefore, we expect a shock to be driven in the Virgo ICM.
For the orbits we derive, the angle of inclination is smaller than the
opening angle of 
the Mach cone ($\simeq 33^\circ$) and we should not expect shock
fronts to be visible on the plane of the sky (\cf\ Rangarajan
\etal\ 1995).
This is illustrated in Figure~\ref{fig:mach}, which shows
the marginally bound orbit with $\ri = 376$~kpc.  Each
circle is drawn centered 
on a point where M86 was at a time $\delta t$ in the past, 
with a radius of $\sound \, \delta t$.  This is
done at equally spaced times to indicate the shape of the Mach cone.
The absence of caustics in Figure~\ref{fig:mach} shows that we would not
see the shock front in projection.  (While this statement is accurate,
except possibly in the small region where the shock front is highly
supersonic, the diagram is schematic.  For example, if M86 were coming
directly towards the Earth, radii of sections of the Mach cone would
be $(1 - \sound^2/v^2)^{-1/2} \simeq 1.20$ larger than drawn.)  Our
lines of sight through the compressed gas behind the shock are
longest where the circles pile up to the southeast of M86, consistent
with the enhancement in X-ray surface brightness seen $\sim3'$
southeast of M86 (Figure~\ref{fig:mach}).

\subsection{Displacement of Stripped Gas}\label{sec:displace}

The prominent plume of gas lying 3\arcmin -- 4\arcmin ($\sim 16$ kpc) north of
M86, noted previously (\eg, Forman et al. 1979), appears to be
physically separated from the dense gas remaining in
the galaxy and it is centered some distance east of the orbits
illustrated in Figure~\ref{fig:m86orb}.  
The removal of a significant fraction of the ISM in a single blob is
expected from rapid ram pressure stripping (Takeda \etal\ 1984).
A highly simplified model for
ram pressure stripping treats the plume as a single particle subject to
gravity and drag due to its motion through the surrounding gas.
Allowing for buoyancy, the net force on the plume due to gravity is
$(\rho - \rhoe) V \gvec$, where $\rho$ is its density, $V$ is its
volume, $\rhoe$ is the density of the ambient ICM, and $\gvec$ is the
local acceleration due to gravity.  The drag force on the plume is
$-\cdrag A \rhoe v \vvec$, where $\cdrag$ is the drag coefficient, $A$
is the plume cross section, $\vvec$ is its velocity, and $v = |\vvec|$.
Thus, the net acceleration of the plume is $\avec = [1 - \rhoe/\rho]
\gvec - [\cdrag\rhoe A / (\rho V)] v \vvec$.  Two of the main
parameters of this model are the density contrast, $\rhoe/\rho$, and
the factor $\cdrag A / V$.  For the density of the ICM, $\rhoe$, we
use the beta model of Schindler \etal\ (1999; $\beta = 0.47$, core
radius $= 2.7'$).  On the grounds that the density of the plume is
likely to be lower now (due to its stripping and ejection from the
confining potential of M86), we set its density contrast at the
projected radius of M86 to be 18 (\cf\ $\sim 16$ from the numbers
above, see \S~\ref{sec:dspec}) and treat
the density of the plume as constant.  The other
factor is determined by treating the plume as a sphere with a constant
radius of 10 kpc, with $\cdrag = 0.75$.  As above, this radius is a
little larger than the observed radius.  To simulate stripping, a gas
blob is placed at the location and velocity of M86 at the time it
is released from the cluster virial radius.  The orbit of the blob is
followed, subject to the gravitational acceleration of the cluster and
the moving galaxy.

This model cannot account for the current
location of the plume if the gravitational potential of M86 is
spherical.  The path
of the blob is determined by the competition between gravity and drag.
If the drag is large, the blob is ejected from M86 early on its orbit,
and slows quickly until it is falling towards the cluster center at
its terminal speed, and ultimately lies farther away from M86 than observed. Its
early ejection also
leaves it too far back along the orbit of M86.  Reducing the relative
significance of the drag causes the blob to be ejected later, bringing
it closer to its observed position
but not far enough from the orbit in the
direction of the cluster center.

This issue can be resolved if the gravitational potential of M86 is
aspherical.  Consider a small gas blob in an aspherical galaxy moving
at an inclined angle through the ICM, illustrated schematically in
Figure~\ref{fig:inclined}.  While ram pressure is insufficient to
eject the blob, it is driven to an equilibrium position where ram
pressure is balanced by gravity.  Since the direction of the
gravitational field must oppose the drag, the external flow must be
perpendicular to the equipotential surface.  The
equilibrium position therefore lies at a location away from the axis
passing through the center of the galaxy and parallel to the flow
(\eg, near the point labeled ``equipotential'' in
Figure~\ref{fig:inclined}).  As the drag starts to overwhelm gravity and
displaces the blob in the direction of the flow, the equipotentials
the blob encounters tilt, so that it is subject to a component of the
gravitational force directed away from the axis.  Because forces
in the direction of the flow are nearly balanced, this off-axis
component of gravity drives the blob
farther away from the axis of the flow.  Thus, a blob stripped from
such a galaxy tends to emerge away from the axis of the flow.
Treating the gas as a fluid, interstellar
gas that is pushed inward by ram pressure along the leading edge of
the galaxy flows out preferentially along the long axis of the
potential, in the direction of the weakest gravitational force.

The model for the aspherical gravitational potential of M86 used here
has the form $\Phi (\rvec) = F (w)$, where $F(w) = \psi \ln [(1 + w /
  c) / (1 + w)] / w$ is a minor modification of the standard NFW form
($\psi \ln [1 / (1 + w)] / w$ in the same notation).  The extra factor
of $1 + w / c$, where $c$ is the concentration parameter, makes the
total mass converge, avoiding the need to truncate the mass
distribution at the virial radius (which would create problems for our
fourth order integrator).  The coordinate $w = \sqrt{(x^2 + y^2)/a^2 +
  z^2/b^2 + 1}$, where $a$ is the NFW
potential scale length and $b$ is chosen to give the desired
ellipticity.  The 
additional 1 under the square root flattens the potential at small
$r$, providing a better model for the gravitational force on an
extended gas blob (of size comparable to $a$) when it is close to the
center of the galaxy.  This modification also makes the integrator
behave better near the center of the galaxy.  The virial radius of M86
was set to 100 kpc and its concentration parameter to 8, roughly the
values expected for a massive galaxy.  The normalizing factor, $\psi$,
was expressed as $\psi = 9 \sigma^2$, so that $\sigma$ is a rough
measure of the line-of-sight velocity dispersion for M86.  The value
used below is $\sigma = 185\ \kms$.  We note that for the model what
matters is the ratio of the drag force on the blob to the binding force
of the M86 potential, allowing for a trade off between the choice of
potential and the density ratio and size of the blob. As a result, the
choice of potential is not very critical.

The two remaining parameters are the axial ratio, $a/b$, and the
orientation of M86.  Consistent with its S0/E3 classification, optical
images show that M86 is highly flattened on scales comparable to the
projected distance of the plume (e.g. $a/b \simeq 1.56$ for the
isophote with $a\simeq 4.5'$ in the deep image of Nulsen \& Carter
1987).  Based on the discussion above, the potential of M86 would need
to have a long axis pointing roughly northward from our line-of-sight.
Therefore, we assume it is oblate, using $a/b = 2$.  The orientation
of M86 is then determined by the
direction of its minor axis.  From the optical images, this has a
position angle of $\simeq 35^\circ$ on the sky.  However, its tilt
with respect to the plane of the sky is unknown.  We have set the
minor axis to point towards us in the north, at an angle of $45^\circ$
from the plane of the sky.  This is roughly
the orientation that maximizes the transverse displacement of the
blob.  Figure~\ref{fig:blob} shows the path of the blob on the sky for
this set of model parameters for the orbit with $\ro$ = 9.1~Mpc and
$\ri$ = 314~kpc.  In this model, the blob is currently slower than M86 by 355~\kms\
along our line of sight and trails it by 38~kpc.

Around the edges of the gas halo in M86, where the shear in the external
flow is strong (Figure~\ref{fig:inclined}), if the effective viscosity
is high, viscous stresses pull the interstellar gas out of the galaxy.
Alternatively, if the viscosity is low, shear instabilities mix the
interstellar gas with the ICM, also stripping it from the galaxy
(Nulsen 1982).  This stripping is aided by the low
pressure due to the Bernoulli effect around the edges of the galaxy, which
tends to pull gas into the path of the flow.  This process works
around all edges of the inclined galaxy, as seen from the
direction of the flow.  However, it is expected to be greatest at the
leading and trailing edges.  The shear in the external flow is expected to
be greatest near the leading edge, favoring stripping there.  The
large ram pressure at the leading edge displaces the edge of the
interstellar gas deeper into the potential of the galaxy, pushing it
towards the trailing edge of the galaxy, where the pressure is lower.
Thus, gas at the trailing edge sits higher in the gravitational
potential of the galaxy, favoring its removal from the galaxy.  This
may account for the apparent double streams of gas seen trailing
M86 and several other galaxies in the composite image.  ``Viscous''
stripping from
the main body of M86 as well as the plume can explain the broad
features of the gas trail, though modeling the finer features, such
as the smaller blobs of gas lying to the east of the galaxy, requires a
  more detailed treatment of the gas dynamics (e.g., numerical
  simulations) and is beyond the scope of this paper.

  The proton
mean free path due to Coulomb collisions is given approximately by
$\lambda \simeq 150 (kT)^2 n_{-3}^{-1}$ pc, where $kT$ is the gas
temperature in keV and the electron density is $10^{-3}
n_{-3}\rm\ cm^{-3}$.  In terms of this, the Reynolds number is
$\reynolds \simeq v L / (s \lambda)$, where $v$ is the flow speed, $s$
is the sound speed, and $L$ is a relevant length scale.  Taking $v =
1550\ \kms$ and $L = 10$ kpc gives $\reynolds \simeq 20$ in the 2.4
keV ICM, but $\reynolds \simeq 3000$ for the gas in the 0.77~keV plume (and
interstellar gas) if it is exposed directly to the external flow.
These values suggest that the external flow can be
largely laminar, while the flow in the cool interstellar gas is
relatively turbulent.  Further complicating matters, the effective
mean free path may be significantly smaller than the Coulomb mean free
path (e.g., Schekochinin \etal\ 2007).  The Reynolds number for this
flow is therefore not well defined, due to the range in gas
temperature and uncertainty in the mean free path.

\section{Summary} \label{sec:summary}

We have argued that the plume and long tail of M86 formed due to ram pressure
stripping forces generated as M86 falls into the Virgo cluster.
Several studies have found a similar interpretation for the
formation of this feature (e.g., Forman et al.\ 1979; Fabian et al.\
1980; Rangarajan et al. 1995, however see Bregman \& Roberts 1990;
Finoguenov et al.\ 2004).
We concentrate on these main results:

\begin{itemize}

\item
the plume and long tail observed in the diffuse emission are created
by ram pressure stripping as M86 falls into
the Virgo cluster.  The tail is 150~kpc in projection (a simple
estimate, which assumes free-fall velocity for M86 and an NFW potential for
M87, gives a lower-limt on the true length of the tail of 380~kpc),
making this the longest ram pressure stripped tail
presently known.

\item
based on the X-ray spectra, we 
detect three distinct components associated with the M86/Virgo cluster
system: the Virgo ICM, with $kT \sim 2.4$~keV; the extended halo of M86,
with $kT \sim 1.2$~keV; and the cooler central and stripped gas of
M86, with $kT \sim 0.8$~keV.  The temperature structure of
the tail is consistent with ram pressure stripping, where the higher entropy
M86 halo gas is stripped first and deposited in the tip of the tail,
and the lower entropy M86 ISM is stripped more recently, constituting the
base of the tail and the plume.

\item
the large line-of-sight velocity of M86, and its position
relative to the Virgo cluster, tightly constrain its orbit,
especially if it is assumed that the gas tail traces the
orbit.  In particular, the observations are inconsistent with a
radial orbit.  We show that M86 is at best only marginally bound
to the Virgo cluster, with an inner turning radius on the order of
300~kpc as expected from its recent in-fall.  Our best-fitting orbital
model requires that M86 be
close to M87, less than 167~kpc closer than or 263~kpc
farther than M87 along our line of sight, which is consistent with the
most recent distance 
estimates based on surface brightness fluctuations (Mei et al.\ 2007)
which give a line-of-sight separation of $0.4\pm0.8$~Mpc.

\item
the prominent plume of gas lying 3\arcmin -- 4\arcmin north of M86
appears to have been rapidly driven from M86 by ram pressure
stripping.  The projected position of the 
plume, which does not lie directly on our best-fit model orbit for M86,
can be understood if M86 has an aspherical potential (as
suggested by optical isophotes).  If M86 moves through the Virgo
ICM at an inclination angle relative to the local flow, the gas at
the trailing edge is more easily stripped, thereby displacing the gas
from the nominal orbit of M86 itself.  This model may also explain the
apparent double streams of gas seen trailing M86, as well as those in
other Virgo galaxies. 

\item
the apparent brightness edge to the southeast seen in {\it ROSAT}
observations is also seen in the {\it Chandra} images.  The edge is well
fit with a two power law gas density model, with an abrupt jump in
density by a factor of $ 1.3^{+0.3}_{-0.4}$ at the edge (consistent
with no jump).  Assuming that this brightness edge
is the shock generated as M86 supersonically falls into the Virgo
cluster, the lack of a well-defined density jump is consistent
with what is expected from our orbital model, which suggests that the
orientation of the Mach cone would make it difficult to detect the
shock edge.

\end{itemize}


\acknowledgments

The financial support for this
work was provided by NASA contracts NAS8-39073, NAS8-38248,
NAS8-01130, NAS8-03060, the Chandra Science Center, and the
Smithsonian Institution.

\clearpage

\clearpage
\begin{deluxetable}{lcccc}
\tablewidth{6.25truein}
\tablecaption{Observation Details \label{tab:obs}}
\tablehead{
\colhead{Obs ID}&
\colhead{Date Obs}&
\colhead{Target}&
\colhead{Active CCDs}&
\colhead{Cleaned Exposure\tablenotemark{a}}\\
\colhead{}&
\colhead{}&
\colhead{}&
\colhead{}&
\colhead{(ksec)}
}
\startdata
318&2000-04-07&M86 &S3, S2, I2, I3&12.964\\
803&2000-05-19&M84 &S3, S2, I2, I3&26.699\\
963&2000-04-07&M86 &S3, S1, S2, I2, I3&13.220\\
1619&2001-06-08&NGC4388 &S3, S2, I2, I3&19.705\\
2882&2002-01-29&NGC4438 &S3, S2, I2, I3&24.891\\
5908&2005-05-01&M84 &S3, S1, S2, I2, I3&36.299\\
5912&2005-03-09&SE of M86 &I0, I1, I2, I3, S2&31.370\\
5913&2005-03-19&M86 Tail Tip &I0, I1, I2, I3, S2&34.744\\
6131&2005-11-07&M84 &S3, S1, S2, I2, I3&37.855\\
\enddata
\tablenotetext{a}{Total cleaned exposure is $\sim$238 ksec.}
\end{deluxetable}

\clearpage
\begin{deluxetable}{lccccc}
\tablewidth{7.0truein}
\tablecaption{Spectral Fits\tablenotemark{a} \label{tab:spectra}}
\tablehead{
\colhead{Region \#}&
\colhead{$kT$}&
\colhead{Abund.}&
\colhead{$\Gamma$}&
\colhead{$\chi^2$/dof}&
\colhead{Net Cnts.}\\
\colhead{}&
\colhead{(keV)}&
\colhead{(solar)}&
\colhead{}&
\colhead{}&
\colhead{}
}
\startdata
1&0.724$^{+0.013}_{-0.014}$//(2.4)&0.42$^{+1.55}_{-0.13}$&1.71$^{+0.51}_{-0.68}$&108/91=1.18&6235\\
1&0.677$^{+0.025}_{-0.021}$/1.291$^{+0.406}_{-0.263}$/(2.4)&0.64$^{+0.56}_{-0.18}$&(1.5)&90/89=1.01&6235\\
2&0.697$^{+0.038}_{-0.033}$/(1.2)/(2.4)&0.31$^{+0.15}_{-0.10}$&&56/57=0.98&3086\\
3&0.653$^{+0.016}_{-0.023}$/(1.2)/(2.4)&0.50$^{+0.90}_{-0.16}$&(1.5)&88/83=1.06&4273\\
4&0.773$^{+0.013}_{-0.013}$/(1.2)/(2.4)&0.47$^{+0.17}_{-0.12}$&&109/84=1.29&8630\\
5&0.854$^{+0.026}_{-0.031}$/(1.2)/(2.4)&0.40$^{+0.19}_{-0.09}$&&94/93=1.01&4461\\
6&0.849$^{+0.064}_{-0.043}$/(1.2)/(2.4)&0.31$^{+0.13}_{-0.10}$&&89/88=1.01&5059\\
7&0.860$^{+0.061}_{-0.048}$/(1.2)/(2.4)&0.43$^{+0.39}_{-0.20}$&&87/74=1.18&2843\\
8&0.785$^{+0.074}_{-0.130}$/(1.2)/(2.4)&0.45$^{+0.42}_{-0.17}$&&44/40=1.09&1749\\
9&0.924$^{+0.116}_{-0.285}$/(1.2)/(2.4)&0.50$^{+2.50}_{-0.27}$&&30/29=1.03&1043\\
10&1.191$^{+0.127}_{-0.134}$/(2.4)\tablenotemark{b}&0.42$^{+0.86}_{-0.25}$&&20/25=0.73&739\\
11&0.865$^{+0.105}_{-0.047}$/1.290$^{+0.240}_{-0.118}$/(2.4)&0.55$^{+0.27}_{-0.15}$&&163/176=0.92&8795\\
12&0.871$^{+0.103}_{-0.025}$/1.176$^{+0.235}_{-0.143}$/(2.4)&0.62$^{+0.28}_{-0.19}$&&240/187=1.29&8914\\
13&0.675$^{+0.111}_{-0.187}$/1.256$^{+0.134}_{-0.232}$/2.723$^{+0.744}_{-0.506}$&0.72$^{+0.48}_{-0.18}$&&343/293=1.17&11033\\
14&0.625$^{+0.008}_{-0.010}$/(1.2)/(2.4)&1.17$^{+0.78}_{-0.10}$&1.82$^{+0.11}_{-0.12}$&459/327=1.40&42002\\
15&0.790$^{+0.055}_{-0.106}$/(1.2)/(2.4)&0.60$^{+0.48}_{-0.18}$&&93/90=1.03&3666\\
16&1.355$^{+0.178}_{-0.065}$/(2.4)\tablenotemark{b}&0.49$^{+0.38}_{-0.17}$&&57/59=0.96&1746\\
17&1.091$^{+0.271}_{-0.159}$/2.422$^{+3.328}_{-0.490}$\tablenotemark{b}&0.40$^{+0.41}_{-0.20}$&&53/45=1.18&3054\\
18&1.085$^{+0.628}_{-0.297}$/2.107$^{+???}_{-0.297}$\tablenotemark{b}\tablenotemark{c}&0.22$^{+0.14}_{-0.12}$&&118/120=0.98&4643\\
19&0.651$^{+0.077}_{-0.143}$/1.065$^{+0.197}_{-0.188}$/(2.4)&0.25$^{+0.21}_{-0.09}$&&147/128=1.15&7325\\
20&1.319$^{+0.276}_{-0.225}$/(2.4)\tablenotemark{b}&0.39$^{+0.78}_{-0.25}$&&30/23=1.29&1087\\
21&1.023$^{+0.043}_{-0.048}$/(2.4)\tablenotemark{b}&0.22$^{+0.10}_{-0.07}$&&147/105=1.39&2912\\
22&0.881$^{+0.065}_{-0.042}$/1.236$^{+0.156}_{-0.101}$/(2.4)&0.32$^{+0.15}_{-0.10}$&&77/75=1.03&2646\\
23&1.186$^{+0.217}_{-0.160}$/2.712$^{+1.003}_{-0.537}$\tablenotemark{b}&0.45$^{+0.34}_{-0.19}$&&214/236=0.91&5193\\
\enddata
\tablenotetext{a}{
  Each region contained up to three thermal APEC components: a low
  temperature, a mid-range temperature, and a high temperature.
  Temperatures are given, in that order, in column~2.  The high
  temperature component models background Virgo ICM emission. Where
  this component could not be accurately measured its temperature was
  fixed at 2.4~keV (see text for discussion).
  Similarly, the mid-range temperature component models M86 halo
  emission, and where the temperature could not be measured it was
  fixed at 1.2~keV.  The abundance for this component was tied to the
  abundance of the cooler fitted component.  A power law 
  component was included when necessary, presumably for unresolved
  point sources. 
  Galactic absorption was assumed throughout.}
\tablenotetext{b}{No low temperature component was included.}
\tablenotetext{c}{Upper bound of 90\% confidence interval could not be
  determined.}
\end{deluxetable}


\clearpage 

\begin{figure}
\includegraphics[width=\linewidth]{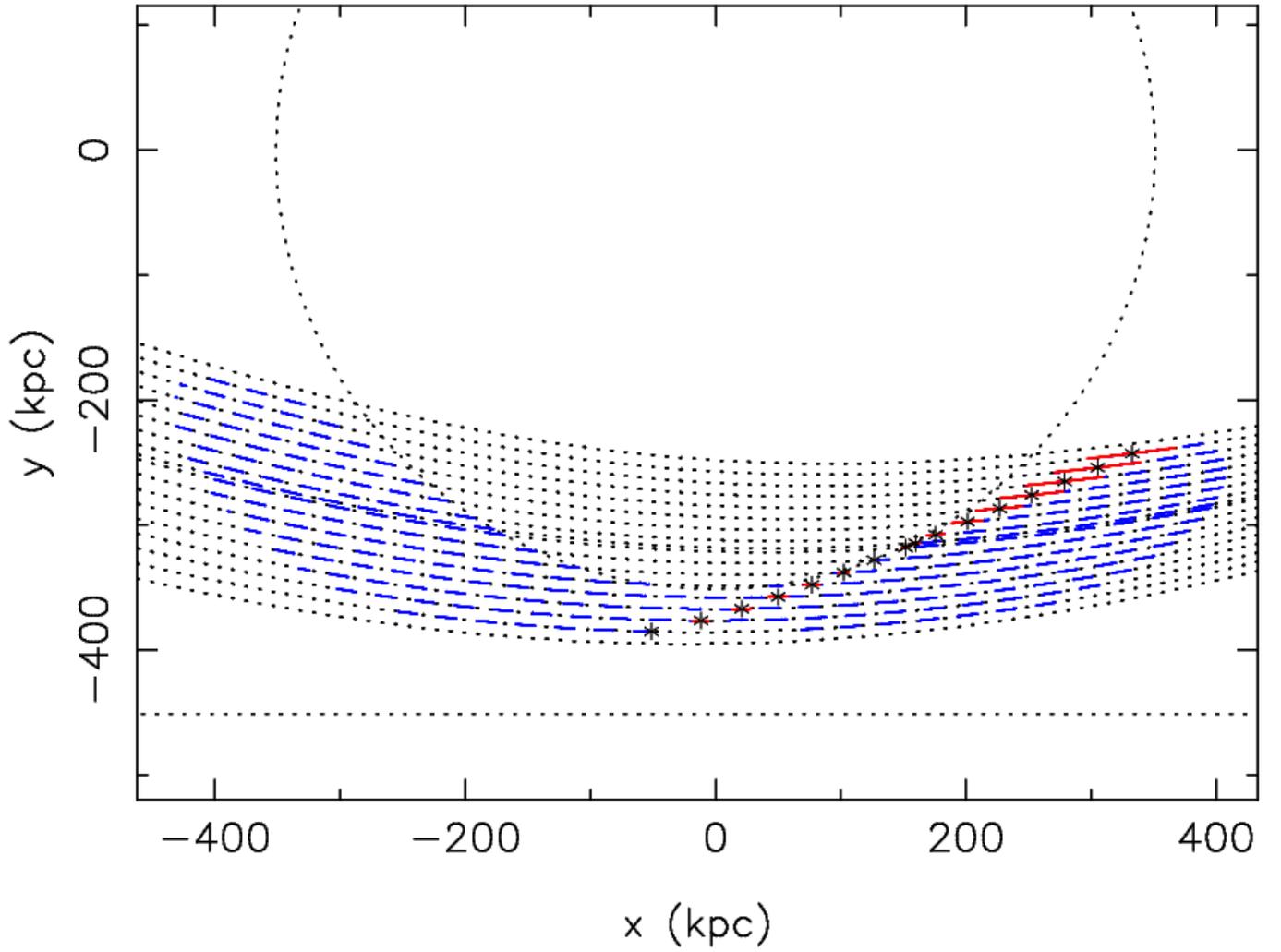}
\caption{Possible marginally bound orbits for M86.  Orbits are plotted
  for inner turning radii in the range 247 -- 395 kpc.  The direction
  of motion is right to left.  The dotted circle shows points 351 kpc
  from the cluster center.  The colored regions (dash-dot and solid lines) show where on each
  orbit the galaxy could be viewed at a projected distance of 351 kpc
  from the cluster center, with a line-of-sight speed of 1550
  \kms\ towards us.  Locations on the orbits where the past orbit
  would also appear to project 100 kpc outside the current position of
  M86 are shown in red (solid lines; the midpoint of each of these regions is
  marked with an asterisk).  Both regions shrink for orbits with smaller
  outer turning radii.  As plotted, all orbits pass through the virial
  radius at $x=1300$ kpc on the $+x$ axis.}
\label{fig:mb}
\end{figure}

\clearpage

\begin{figure}
\includegraphics[width=\linewidth]{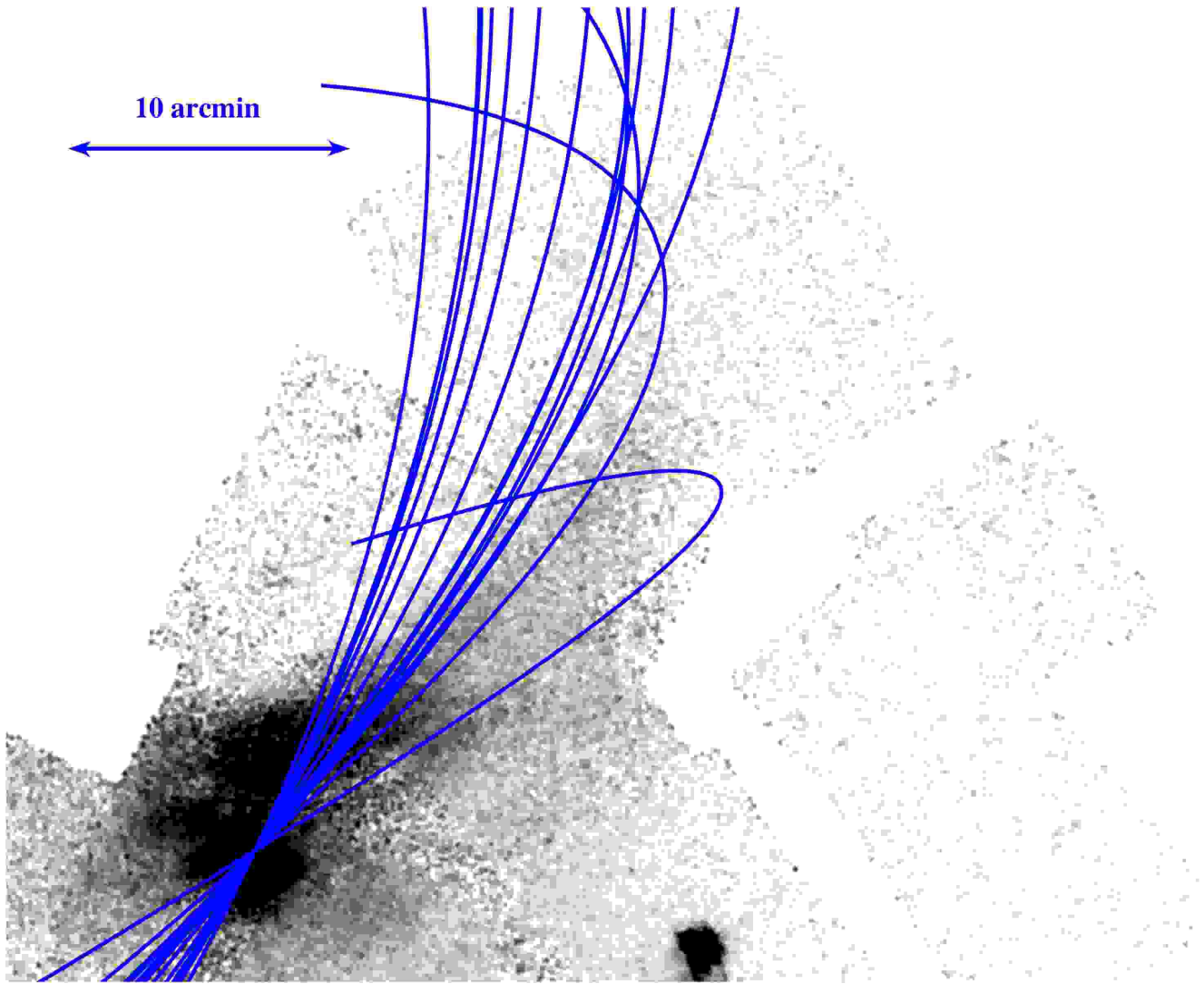}
\caption{Candidate marginally bound orbits for M86.  
  An  orbit is
  shown projected onto the smoothed 0.5 -- 2 keV image of M86 for 
  each of the
  midpoints of the red regions shown in Figure~\ref{fig:mb}
  (marked with an asterisk).
  Orbits are
  plotted from the point where they cross the virial radius (sometimes
  outside the field shown).}
\label{fig:projmb}
\end{figure}

\clearpage

\begin{figure}
\includegraphics[width=\linewidth]{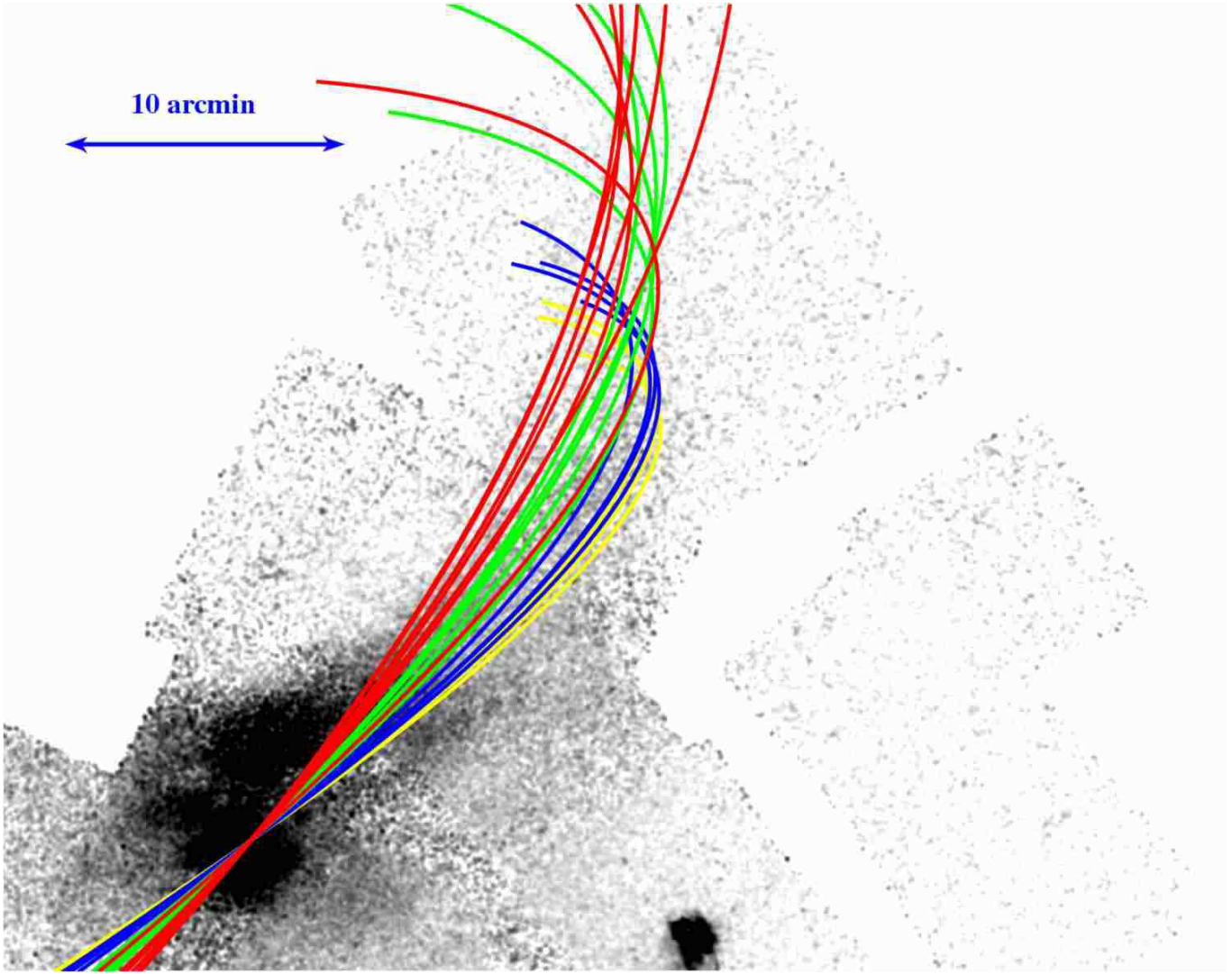}
\caption{Possible orbits for M86.  As in Figure~\ref{fig:projmb}, one
  representative orbit is shown projected onto the smoothed 0.5 -- 2
  keV image of M86 for
  the remaining plausible orbits.  Colors indicate the outer turning
  radius as follows: $\ro = 8.76$ Mpc, yellow; $\ro = 9.11$ Mpc, blue;
  $\ro = 19.3$ Mpc, green; $\ro = \infty$, red.}
\label{fig:m86orb}
\end{figure}

\clearpage

\begin{figure}
\includegraphics[width=\linewidth]{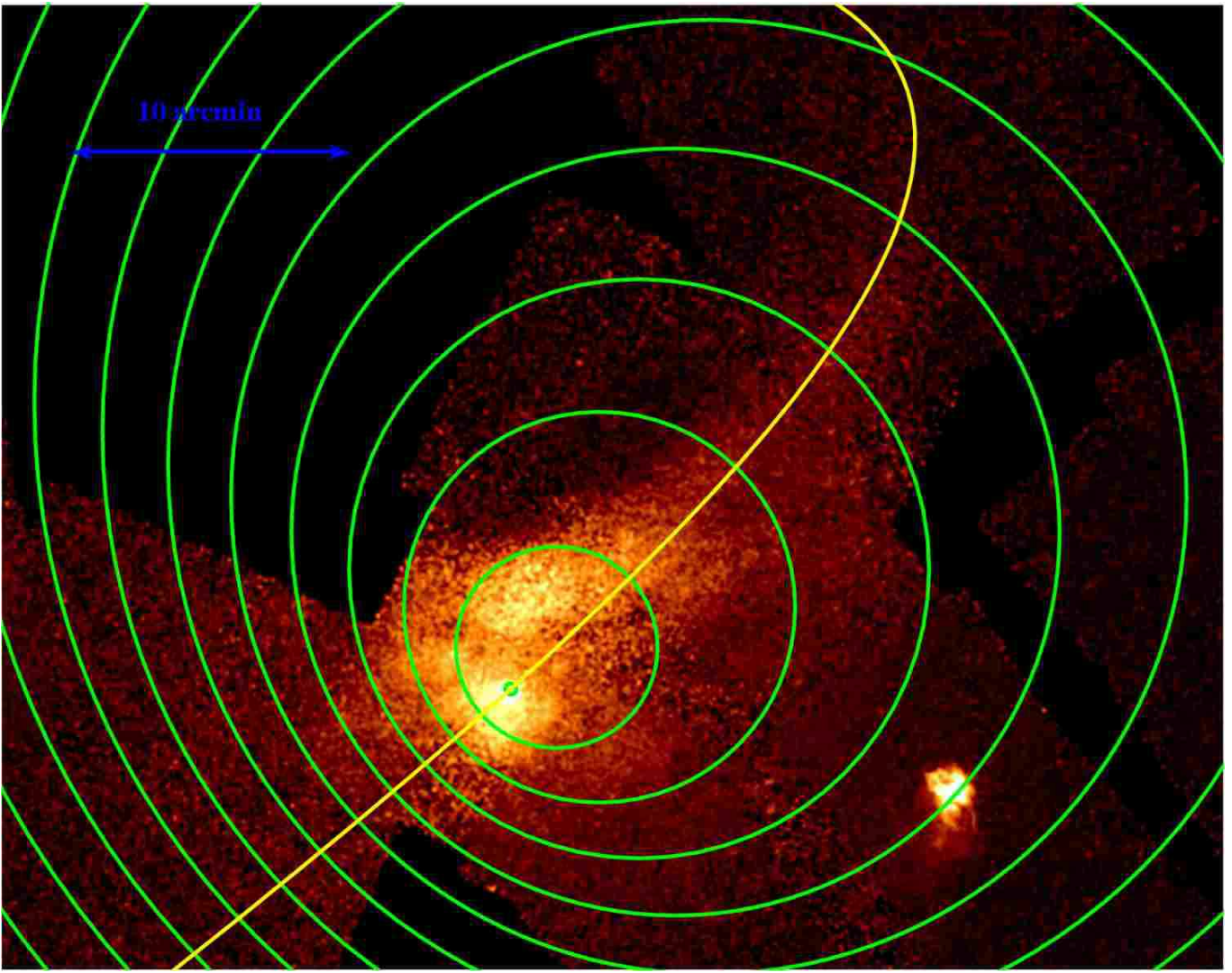}
\caption{Illustration of the Mach cone for M86.  Shown in yellow is
  the marginally bound orbit with $\ri = 376$ kpc, one of the better
  candidates for the orbit of M86.  Centered on the orbit, at equally
  spaced times, circles are plotted on the plane of the sky with radii
  $c_{\rm s} \, \delta t$, where $\delta t$ is the time before the present
  when M86 was at that location and $c_{\rm s}$ is the sound speed
  ($850\rm\ km\ s^{-1}$).}
\label{fig:mach}
\end{figure}

\clearpage
\begin{figure}
\includegraphics[width=0.9\linewidth]{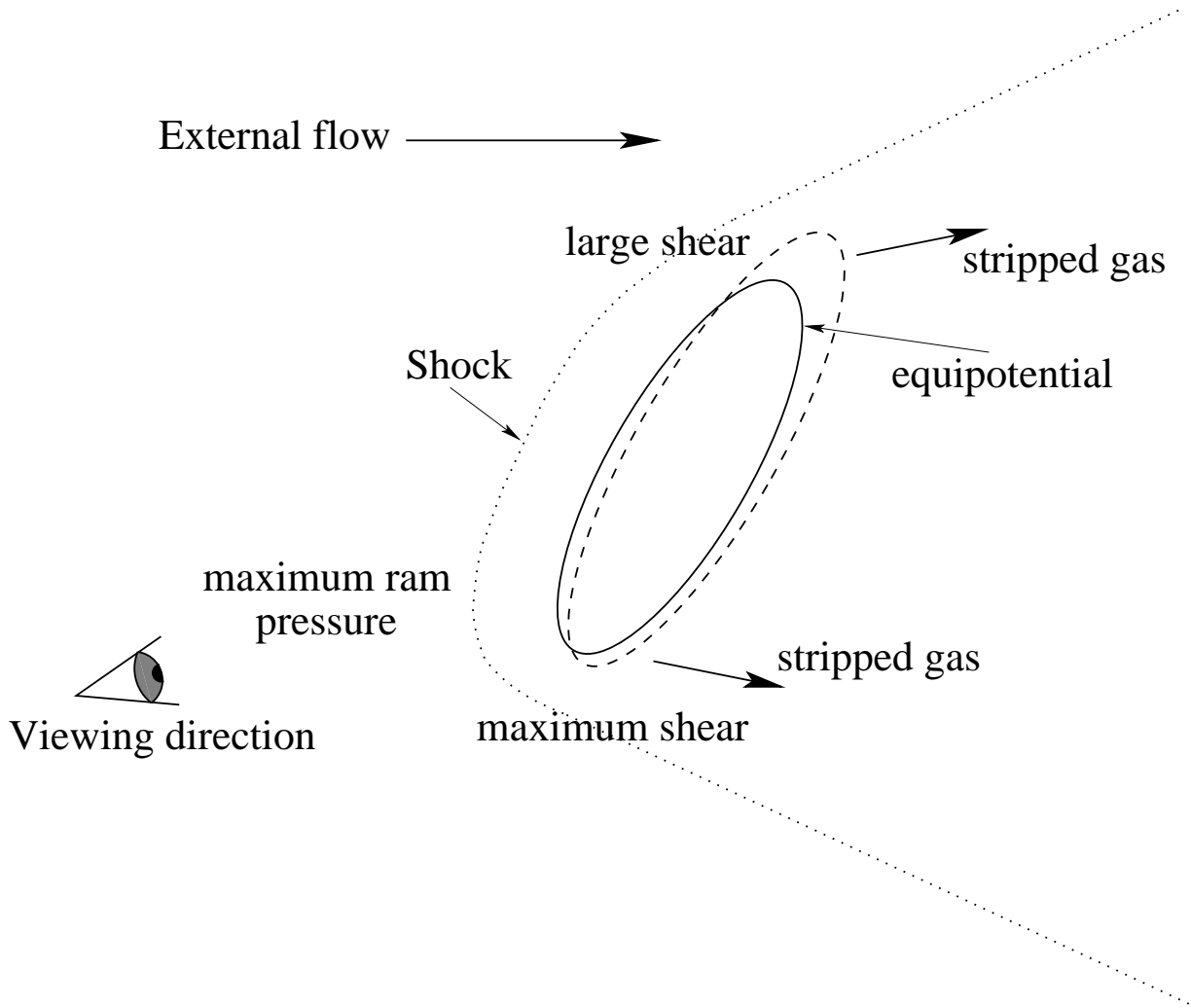}
\caption{Stripping from an aspherical galaxy.  Schematic diagram of an
  inclined, aspherical galaxy moving supersonically to the left
  through an external medium.  The solid ellipse represents an
  equipotential of the galaxy, the dotted line represents the shock
  front and the dashed line represents the outer boundary of the
  interstellar gas.  The ram pressure is maximized near the leading
  edge of the galaxy, where the shock is normal to the external gas
  flow, and lower on inclined sections of the shock front.  The gas
  flow is fastest around the edges of the galaxy, causing local minima
  in pressure that pull gas out sideways from the galaxy.  Gas
  displaced by ram pressure from the leading edge of the galaxy piles
  up along the trailing side and edge.}
\label{fig:inclined}
\end{figure}

\clearpage

\begin{figure}
\includegraphics[width=\linewidth]{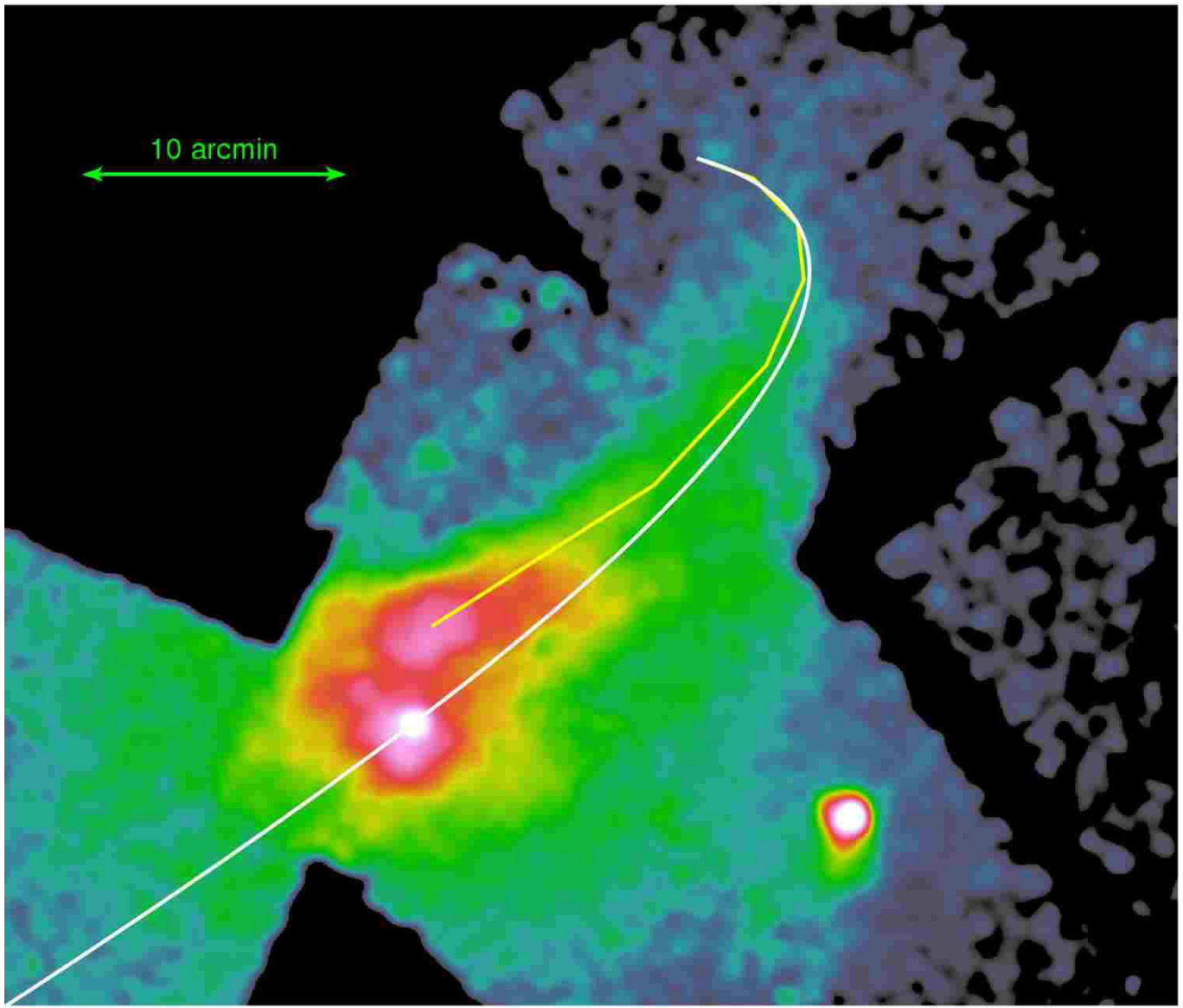}
\caption{Model orbit for ram pressure stripped blob.  The orbit of M86
  is shown in white and the (coarsely sampled) orbit of the blob is
  shown in yellow.  Note that the path of the blob follows a trail of
  denser gas.}
\label{fig:blob}
\end{figure}

\clearpage
\appendix
\section{Constraining M86's Orbit} \label{ap:orbit}

In cylindrical polar coordinates, with the origin at the cluster
center and the orbit (instantaneously) in the $x$ -- $y$ plane, the
position of M86 can be expressed as $\rvec = r \erad$ and its velocity
as $\vvec = \vrad \erad + \vphi \ephi$, where $\erad$, $\ephi$, $\ez$
are the coordinate basis vectors.  When M86 is viewed along the unit
vector, $\nhat$, its location projected onto the plane of the sky is
$\rvec - \nhat \cdot \rvec \nhat$.  Expressing $\nhat$ in terms of the
coordinate vectors as $\nhat = a' \erad + b' \ephi + c' \ez$, the
projected separation, $s$, of M86 from the cluster center is given by
$s^2 = (\rvec - \nhat \cdot \rvec \nhat)^2 = r^2 (1 - a'^2)$, which
requires $a' = \pm \sqrt{1 - s^2/r^2}$ (and $r > s$).  Since $\nhat$ is
a unit vector, $a'^2 + b'^2 + c'^2 = 1$, further requiring $b' =
\pm\sqrt{s^2/r^2 - c'^2}$.  The remaining parameter, $c'$, is the cosine
of the polar angle of the viewing direction (\ie, the orbit
inclination).  With the two values of $a'$ above, the parametrization
$(b', c') = (\cos\xi, \sin\xi) s/r$, for $0 \le \xi < 2\pi$, specifies
the complete range of viewing directions that project M86 at the
observed distance from the cluster center.  In this notation, the
velocity of M86 along our line-of sight is $\vlos = \nhat \cdot \vvec
= a' \vrad + b' \vphi$.  For radial orbits with M86 inbound ($\vrad < 0$), $\vlos = a \vrad = \sqrt{1 - s^2 / r^2} \vrad < 0$.  This
vanishes for $r \to s$ (where a radial orbit must be perpendicular to
our line-of-sight), while the radial speed, $|\vrad|$, decreases with
increasing $r$, so that $|\vlos|$ is maximized for some $r > s$.

The inner and outer turning radii, $\ri$ and $\ro$, respectively, are related to the specific energy of an orbit by $E = (\ro^2 \phio -
\ri^2 \phii) / (\ri^2 - \ro^2)$, where $\phii = \phi(\ri)$ is the
gravitational potential at $\ri$ and $\phio$ is that at $\ro$.  The
specific angular momentum of the orbit, $\ell$, is given by $\ell^2 =
2 \ri^2 \ro^2 (\phio - \phii) / (\ro^2 - \ri^2)$.  If M86 is bound to
the Virgo cluster, then, since $r \ge \ri$ and its speed, $v \ge
|\vlos|$, $0 > E = v^2/2 + \phi(r) \ge \vlos^2/2 + \phii$, which
requires $\phii < - \vlos^2/2$.  Taking $\vlos = -1550\ \kms$, this
gives an upper limit on the inner turning radius of $\ri < 489$ kpc.
To obtain a lower limit on $\ro$, note that $\partial E/\partial\ro =
(\vko^2 - \vo^2) \ro / (\ro^2 - \ri^2)$, where the Kepler speed at
$\ro$ is given by $\vko^2 = \left. r\, d\phi/dr \right|_{r = \ro}$ and
the speed at the outer turning radius is $\vo^2 = \ell^2/\ro^2 = 2 (E
- \phio)$.  The net acceleration at the outer turning radius is
inward, requiring $\vo < \vko$, so that $E$ is an increasing function
of $\ro$.  Interchanging $\ri$ and $\ro$ in this argument and noting
that the net acceleration is outward at the inner turning radius
($\vki < \vi$) shows that $E$ is also an increasing function of $\ri$.
Since $dE = \partial E / \partial \ri \, d\ri + \partial E/\partial
\ro \, d\ro$, for a fixed energy, maximizing $\ri$ minimizes $\ro$.
For M86, $v \ge |\vlos|$ and $r \ge s$, so that its energy $E \ge
\emin = \vlos^2 / 2 + \phi(s)$.  To obtain $E = \emin$ with $\ri = s$
requires $\ro = \romin \simeq 3.5$ Mpc, or $\simeq 2.7$ times the
virial radius of the cluster.  For values of $\ri$ larger than $s$,
the total energy would need to match or exceed $\vlos^2 / 2 + \phii >
\emin$.  The value of $\ro$ required to
attain this energy is an increasing function of $\ri$, unless the mean
density of the cluster increases with radius.  Thus, $\romin$ provides
a lower limit on the outer turning radius of M86's orbit.

At a fixed point on an orbit, the requirement $\vlos = a' \vrad + b'
\vphi$ determines the possible viewing directions, if any, that could
model M86.  For each of the two solutions $a' = \pm\sqrt{1 - s^2/r^2}$,
the requirement $\vlos = a' \vrad + b' \vphi$ gives $b'
= (\vlos - a' \vrad) 
/ \vphi$.  If either of these gives $b'$ in the range $[-s/r, s/r]$,
then it corresponds to two possible viewing directions, with $c' = \pm
\sqrt{s^2/r^2 - b'^2}$ (symmetrically placed above and below the plane
of the orbit).  Thus, for each point on an orbit, there can be zero,
two, or four potential viewing directions that would place M86 at the
observed separation from the cluster center and give it the observed
line-of-sight velocity.  Figure~\ref{fig:mb} shows a range of marginally
bound orbits ($\ro = \infty$), with regions that meet these conditions
marked in color.  Since $|\vlos| < \vlosmax = \sqrt{1 - s^2/r^2}
|\vrad| + s/r |\vphi|$, which is an increasing function of $\ro$ for
fixed $r$ and $\ri$, the range of an orbit where these conditions are
met shrinks as $\ro$ decreases, \ie, as the orbit becomes more
tightly bound.

If the trail of stripped gas lies along the orbit of M86, then the
past orbit must project away from the cluster center at least as far
as the gas tail extends.  From above, $(\rvec - \nhat \cdot \rvec
\nhat) / s$ is a unit vector in the plane of the sky that points from
the cluster center towards M86.  For each potential location of M86 on
an orbit, we have computed $d(\rvec') = \rvec' \cdot (\rvec - \nhat
\cdot \rvec \nhat) / s$ for values of $\rvec'$ ranging over the past
orbit.  If stripping is to account for the location of the gas tail,
$d(\rvec')$ must increase initially as $\rvec'$ moves backward along
the orbit and the maximum value of $d(\rvec') - s$ must exceed the
distance, $\sim 100$ kpc, that the tail projects beyond M86 in the
direction away from the cluster center.  Locations on the orbits where
the line-of-sight speed can attain $-1550\ \kms$ and these conditions
are also met are shown in red in Figure~\ref{fig:mb}.

\end{document}